\documentclass[useAMS,usenatbib]{mn2e}

\usepackage{graphicx}
\usepackage{aas_macros}
\usepackage{amsmath}
\usepackage{amssymb}
\newcommand{\z}[1]{$z\sim#1$} 
\newcommand{\R}{$\cal R$}
\newcommand{\Un}{$U_n$}

\newcommand{\chandra}{\textit{Chandra}}

\newcommand{\lala}{LALA CETUS}
\newcommand{\lynx}{Lynx}
\newcommand{\egs}{EGS1}
\newcommand{\gws}{GWS}
\newcommand{\hdfn}{HDF-N}
\newcommand{\ergscm}{$\mathrm{erg\;s^{-1} cm^{-2}}$}
\newcommand{\ergs}{$\mathrm{erg\;s^{-1}}$}

\newcommand{\dd}{\mathrm{d}}
\newcommand{\Lx}{L_\mathrm{X}}
\newcommand{\fx}{f_\mathrm{X}}
\newcommand{\giv}{\;|\;}
\newcommand{\amend}[1]{{#1}} 

\begin{document}

\date{Accepted 2009 October 05. Received 2009 September 03; in original form 2009 April 29}

\pagerange{\pageref{firstpage}--\pageref{lastpage}} \pubyear{2009}

\title{The evolution of the hard X-ray luminosity function of AGN}

\author[J. Aird et al.]{J. Aird$^1$\thanks{jaird@ucsd.edu}\thanks{Present address: Center for Astrophysics and Space Sciences (CASS), Department of Physics, University of California, San Diego, CA 92093, USA}, K. Nandra$^1$, E. S. Laird$^1$, 
	A. Georgakakis$^2$,
	M. L. N. Ashby$^3$,
	P. Barmby$^4$,
	\newauthor
	A. L. Coil$^5$,
	J.-S. Huang$^3$,
	A. M. Koekemoer$^6$,
	C. C. Steidel$^7$,
	and
	C. N. A. Willmer$^8$\\
$^1$Astrophysics Group, Imperial College London, Blackett Laboratory, Prince Consort Road, London SW7 2AZ, United Kingdom\\
$^2$National Observatory of Athens, Institute of Astronomy, V. Paulou \& I. Metaxa, Athens 15236, Greece\\
$^3$Harvard-Smithsonian Center for Astrophysics, 60 Garden Street, Cambridge, MA 02138, USA\\
$^4$Department of Physics \& Astronomy, University of Western Ontario, London, ON N6A 3K7, Canada\\
$^5$Center for Astrophysics and Space Sciences (CASS), Department of Physics, University of California, San Diego, CA 92093, USA\\
$^6$Space Telescope Science Institute, 3700 San Martin Drive, Baltimore MD 21218, USA\\
$^7$California Institute of Technology, MS 105-24, Pasadena, CA 91125, USA\\
$^8$Steward Observatory, University of Arizona, Tuscon, AZ 85721, USA
}

\maketitle

\label{firstpage}

\begin{abstract}
We present new observational determinations of the evolution of the 2--10 keV X-ray luminosity function (XLF) of AGN. We utilise data from a number of surveys including both the 2 Ms \chandra\ Deep Fields and the AEGIS-X 200 ks survey, enabling accurate measurements of the evolution of the faint end of the XLF. We combine direct, hard X-ray selection and spectroscopic follow-up or photometric redshift estimates at $z< 1.2$ with a rest-frame UV colour pre-selection approach at higher redshifts to avoid biases associated with catastrophic failure of the photometric redshifts. Only robust optical counterparts to X-ray sources are considered using a likelihood ratio matching technique. A Bayesian methodology is developed that considers redshift probability distributions, incorporates selection functions for our high redshift samples, and allows robust comparison of different evolutionary models. 
We statistically account for X-ray sources without optical counterparts to correct for incompleteness in our samples. We also account for Poissonian effects on the X-ray flux estimates and sensitivities and thus correct for Eddingtion bias.
We find that the XLF retains the same shape at all redshifts, but undergoes strong luminosity evolution out to $z\sim 1$, and an overall negative density evolution with increasing redshift, which thus dominates the evolution at earlier times. We do not find evidence that a Luminosity-Dependent Density Evolution, and the associated flattening of the faint-end slope, is required to describe the evolution of the XLF. We find significantly higher space densities of low-luminosity, high-redshift AGN than in prior studies, and a smaller shift in the peak of the number density to lower redshifts with decreasing luminosity. The total luminosity density of AGN peaks at $z=1.2\pm0.1$, but there is a mild decline to higher redshifts. We find $> 50$\% of black hole growth takes place at $z>1$, with around half in $L_\mathrm{X}<10^{44}$ erg s$^{-1}$ AGN.

\end{abstract}

\begin{keywords}
galaxies: active -- X-rays: galaxies -- galaxies: evolution -- galaxies: luminosity function, mass function.
\end{keywords}

\section{Introduction}
\label{sec:intro}

Active Galactic Nuclei (AGN) are an important constituent of the Universe, playing a vital role in the formation and evolution of galaxies and having a wider influence on the structure of the Universe. Determining the distribution and evolution of AGN accretion activity throughout the history of the Universe, traced by the luminosity function, is essential to constrain models of super-massive black-hole formation and growth, the triggering and fuelling of AGN and their co-evolution with galaxies. This requires large samples of objects spanning a wide range of redshifts and luminosities to accurately determine the shape of the luminosity function and how this evolves over cosmic time. 

X-ray surveys provide a highly efficient method of selecting AGN, including unobscured and moderately obscured sources and low-luminosity sources that may not be identified as AGN at optical wavelengths. X-ray emission is relatively unaffected by absorption, particularly at hard X-ray energies ($\gtrsim 2$ keV) for moderate column densities ($\mathrm{N_H}\lesssim 10^{23}$ cm$^{-2}$), and thus provides a direct probe of the accretion activity. 
Large efforts have been made to obtain follow-up observations of X-ray sources in various X-ray surveys to determine their redshifts, and thus allow the X-ray luminosity function to be determined \citep[e.g.][]{Barger03c,Szokoly04,Trouille08}.
Such studies of the X-ray luminosity function (XLF) have revealed AGN are a strongly evolving population, with the overall distribution shifting to lower luminosities between $z\sim1$ and the present day. This evolution can be described by a Pure Luminosity Evolution parameterisation \citep[PLE,][]{Boyle93,Barger05} in which the XLF retains the same shape (a double power-law with a break at a characteristic luminosity, $L_*$), but shifts to lower luminosities as redshift decreases. However, recent data indicate that a more complex, Luminosity Dependent Density Evolution (LDDE) parameterisation is required to describe the evolution of the XLF at both low ($z\lesssim1$) and high ($z\approx 1-4$) redshifts \citep[e.g.][]{Ueda03,LaFranca05,Silverman08,Ebrero09,Yencho09}. In this scheme the shape of the XLF changes with redshift, with the faint-end slope flattening as redshift increases. This evolution is also characterised by a shift in the peak of the space density of AGN towards lower redshifts for lower luminosities.

However, there are remaining uncertainties as to the exact form of the evolution, especially at high redshifts. In part, this is due to the difficulties of accurately measuring the faint end of the XLF, where incomplete and uncertain redshift information is a serious issue. The \chandra\ Deep Fields (CDF-N: \citealt{Alexander03}; CDF-S: \citealt{Giacconi02}) have been key in studies of the XLF, providing the deepest available X-ray data and substantial programmes of spectroscopic follow-up of the X-ray sources to determine redshifts. However, even in these fields a high fraction ($\sim 50$\%) of the X-ray sources remain spectroscopically unidentified, mainly because the optical counterparts of the faintest X-ray sources are also faint ($R\gtrsim24$), and thus beyond the limits of current instrumentation. This can severely bias determinations of the XLF, especially at high redshifts \citep[e.g.][]{Barger05}.

One general approach to address this issue is to determine photometric redshifts for the X-ray sources that are too faint for optical spectroscopic follow-up. This significantly improves the completeness of the samples, although many X-ray sources which lack detectable optical counterparts remain unidentified. However, there are considerable uncertainties in such redshift estimates, particularly for the faintest optical counterparts, and a risk of catastrophic failures in the redshift determinations, which is particularly acute for sources at $z\sim 1-3$. It is also necessary to correct for remaining incompleteness in such samples, which can prove difficult. Additionally, there is a significant risk of associating the incorrect optical counterpart with an X-ray source, due to the high surface density of optical sources at these faint magnitudes, and thus an X-ray source may be assigned completely incorrect redshift information. For example, \citet{Barger03c} were able to assign optical counterparts to $\sim 85$\% of the X-ray sources in the CDF-N, but noted that $\sim 25$\% of the optical identifications at $R=24-26$ are expected to be spurious. Such false identifications of faint X-ray sources could significantly bias determinations of the XLF. 
To reduce these effects many authors  \citep[e.g.][]{Ueda03,Barger05,Silverman08} set relatively high X-ray flux limits, thus increasing the spectroscopically identified fraction and reducing the numbers of optically-faint or unidentified sources, but this also reduces the sensitivity to the key low-luminosity AGN. \amend{The small number of fields with sufficiently high spectroscopic completeness also means most results are susceptible to cosmic variance.}

An alternative approach to the incompleteness issue was adopted by \citet{Aird08} \citep[see also][]{Nandra05} to determine the XLF at high redshift, where the effects can be most severe. The Lyman-break technique \citep[e.g.][]{Steidel93,Steidel03} was used to pre-select a sample of objects at \z3 based on their broadband colours in three filters, and deep X-ray observations were used to identify AGN. This technique provides an incomplete sample \amend{(missing heavily reddened sources in particular)}, but the optical selection functions are well-defined and were derived using simulations of the optical data. This allowed corrections for the incompleteness to be applied, and thus the faint end of the XLF at \z3 could be accurately measured. The number density of low-luminosity AGN was at least as high as previous estimates, and there was no evidence for the flattening of the faint-end slope at $z>1$ that is characteristic of LDDE parameterisations. However, this work did not account for the uncertainties in the redshift estimates for sources without spectroscopic confirmation, and did not present an evolutionary model.

In addition to uncertainties in the optical follow-up and redshift estimates, a number of issues arising from the X-ray observations may also affect the determination of the evolution of the XLF. The faintest sources in \chandra\ surveys are detected with very small numbers of counts, and thus Poissonian effects can be significant. Derived X-ray fluxes can thus be highly uncertain, and may be underestimated by up to 50\% due to the Eddington bias \citep{Laird09}, leading to uncertainties in the luminosity. \citet{Georgakakis08} presented a new method of determining the sensitivity of \chandra\ observations, which fully accounts for the Poissonian nature of the detection limits, and accounted for the flux uncertainties and Eddington bias in the determination of the X-ray number counts. However, these techniques have not been incorporated in previous studies of the XLF.

In this paper we present a new study of the evolution of the XLF, in which we incorporate new deep X-ray observations and develop improved methodologies to address some of the issues regarding optical counterparts, completeness, photometric redshifts and X-ray fluxes described above. Section \ref{sec:data} describes our data, including our samples of hard X-ray selected AGN and the data used to perform high-redshift colour pre-selection of AGN. 
We adopt a likelihood ratio method to ensure we assign secure optical counterparts to the faintest X-ray sources. 
In section \ref{sec:bayes_xlf} we develop a Bayesian methodology to determine the evolution of the XLF, which accounts for uncertainties in redshift estimates and X-ray fluxes, incorporates the improved sensitivity calculations of \citet{Georgakakis08}, corrects for remaining incompleteness of the optical identifications and allows a robust comparison of different evolutionary models. Section \ref{sec:highz} describes how high-redshift rest-frame UV colour-selected samples are incorporated into this methodology \citep[building on the work of][]{Aird08}, to provide a more robust determination of the evolution at $z\gtrsim1.2$.  Our results are presented in section \ref{sec:xlf_evol}, and discussed further in section \ref{sec:discuss}.

A flat cosmology with $\Omega_\Lambda=0.7$ and $h=0.7$ is adopted throughout.

\section{Data and samples}
\label{sec:data}
\label{sec:xsamples}

To determine the evolution of the X-ray luminosity function requires large samples of AGN spanning a wide range of luminosities and redshifts. For our initial investigation of the evolution of the XLF
our approach is similar to most previous studies of the XLF \citep[e.g.][]{Ueda03,Barger05,Silverman08,Ebrero09} in that we attempt to assign redshifts to all sources in a purely X-ray selected catalogue, utilising photometric redshifts to improve the completeness. Thus AGN are detected over a wide redshift range, which is essential for probing the evolution. Our methodology (see section \ref{sec:bayes_xlf}) \amend{requires probability distributions that describe the uncertainties in the photometric redshift estimates. We therefore adopt the best available photometric redshifts with suitable probability distributions, or generate our own from the available data}. To obtain the best constraints on the XLF a high fraction of spectroscopic identifications is also required. Additionally, we require hard X-ray coverage to reduce biases against moderately absorbed sources. Three deep \chandra\ surveys have the necessary spectroscopic follow-up, combined with the deep multiwavelength imaging required for photometric redshifts: the two \chandra\ Deep Fields and the AEGIS-X survey. We supplement this sample with two larger area \textit{ASCA} surveys with near complete spectroscopic follow-up, which are needed to constrain the bright end of the XLF. Table \ref{tab:hardx} summarises the numbers of sources and fraction with optical counterparts and spectroscopic redshifts. Further details and a description of the photometric redshift estimates are given in section {\ref{sec:hardxdata} below.
 
For the high-redshift colour preselection approach the requirements are different. Deep X-ray data are again required, but the soft (0.5--2 keV) band, where the sensitivity of \chandra\ is highest, may be used for selection as this probes relatively hard \emph{rest-frame} energies at $z\gtrsim2$, and thus should not be severely biased by absorption effects. Deep optical imaging data in \Un G\R, or a comparable filter set are required for the colour pre-selection, with deep data at $U$-band wavelengths being the major requirement. A description of the data for the high-redshift work is given in section \ref{sec:highzdata}.

\subsection{Hard X-ray selected samples of AGN}
\label{sec:hardxdata}
 \begin{table*}
 \caption[Summary of hard X-ray selected samples]{
Details of the hard X-ray selected samples/fields.
}
\begin{center}
\begin{tabular}{c c c c c c c c}
\hline
Field		& RA		& Dec		& X-ray exposure & Survey area$^{a}$ & $N_\mathrm{X}$$^{b}$ & $F_\mathrm{opt}$$^{c}$ & $F_\mathrm{spec}$$^{d}$\\
                   & (J2000)		& (J2000)		& (ks)		    & (arcmin$^2$) &                              &                                   &         \\
\hline	
CDF-S	& 03:32:28 	& $-$27:48:30	         &  1933.4 		    & 436.2      & 248                      &  68\%                       &  44\% \\
CDF-N      & 12:36:55 & +62:14:18             &   1862.9			    & 436.0      & 303                      &  67\%                       &  55\% \\
AEGIS-X   & 14:17:43 &  +52:28:25       	&   188.7                         &  $1\;155.6$  & 397                      &  86\%                       &  40\% \\
ALSS         & 13:14:00       & +31:30:00	&   ...		   &  $20\;880.0$     &  34                        &  97\%                       &  97\% \\ 
AMSS        &  ...	& ...		&  ...			   & $294\;386.0$    &  109                     &  98\%                       & 98\% \\
\hline
\end{tabular}
\end{center}
\begin{flushleft}
$^{a}$Area covered by both X-ray data and required optical imaging.\\
$^{b}$Total number of hard X-ray selected sources in the area covered.\\
$^{c}$Fraction of hard X-ray sources with secure optical counterparts.\\
$^{d}$Fraction of hard X-ray sources with secure optical counterparts and spectroscopic redshifts.\\
\end{flushleft}
\label{tab:hardx}
\end{table*}

\subsubsection{\chandra\ Deep Field-South}
\label{sec:cdfs}

The original $\sim 1$ Ms observation of the \chandra\ Deep Field South \citep[CDF-S,][]{Giacconi02} has recently been supplemented with additional Director's Discretionary Time to provide a total of 
$\sim 2$ Ms exposure \citep{Luo08}, which we use in this paper. Data reduction and source detection was carried out using the procedures described in \citet{Laird09}. For our investigation of the XLF we use the hard (2--7 keV) selected source catalogue of objects detected with false Poisson probability $< 4 \times 10^{-6}$, which provides a sample of 248 X-ray sources. The X-ray sensitivity (area curve) is determined using the procedure described in \citet{Georgakakis08}.

Deep optical data have been obtained in several bands in the CDF-S using both ground-based observatories and the \textit{Hubble} Space Telescope. We have searched for optical counterparts using the COMBO-17 survey \citep{Wolf04} source catalogue, in which sources are detected in deep $R$-band images (10$\sigma$ limiting AB magnitude $\sim 25.6$) obtained with the WFI instrument on the ESO 2.2m telescope at La Silla. Counterparts were found using the implementation of the likelihood ratio method \citep{Sutherland92,Ciliegi03} described in \citet{Laird09}. This method allows us to assign secure optical counterparts to the X-ray source, accounting for the X-ray and optical positional uncertainties, the a priori probability of a counterpart with a given magnitude, and the probability of a spurious optical counterpart. We restrict our sample to objects with likelihood ratios $>0.5$, which provides optical counterparts for 169 ($\approx 68$\%) of the hard-band selected X-ray sources. 

A number of spectroscopic programmes have been carried out in the CDF-S. Following \citet{Luo08}, we have combined spectroscopic surveys from \citet{Szokoly04}, \citet{LeFevre04}, \citet{Mignoli05}, \citet{Ravikumar07}, \citet{Popesso08} and \citet{Vanzella08}. This provides spectroscopic redshifts for 110 sources (65\% of the optical counterparts, 44\% of the entire hard X-ray selected sample).

A variety of photometric redshift estimates have been published in the CDF-S area, focussed on both the X-ray source population \citep[e.g.][]{Zheng04} and larger populations of optically selected sources \citep[e.g.][]{Wolf04,Mobasher04}. However, we have chosen to use our own photometric redshifts estimates using the Bayesian photometric redshift code \citep[BPZ,][]{Ben00}, which allows us to use the full redshift probability distribution output by the BPZ code, and thus account for the significant uncertainties in photo-$z$ estimates in our determination of the XLF (see section \ref{sec:bayes_xlf} below). Such probability distributions are not available for the published catalogues. 

Our photo-$z$ estimates were determined using the full 17-band photometry for the X-ray sources (with secure optical counterparts) from the COMBO-17 catalogues \citep{Wolf04}. The BPZ code uses a $\chi^2$ minimisation template fitting approach. A set of 6 template spectra are used, consisting of the four standard templates from \citet{Coleman80} (E/S0, Sbc, Scd, and Irr) and two star-bursting galaxy templates from \citet{Kinney96}. Additional points in colour space are calculated by interpolating between the templates. The BPZ code also adopts a Bayesian approach, applying priors based on the expected fractions and redshift distributions for different templates, and marginalising over the templates to produce the final redshift probability distribution, $p(z)$. These distributions describe the uncertainties in the redshift estimate due to both photometric errors and colour, redshift and template degeneracies, which often result in multi-modal probability distributions. We note that BPZ does not include AGN or QSO templates. Broad-line QSOs are likely to be optically bright, and may be more easily identified spectroscopically due to their clear spectroscopic features. The remaining AGN population is expected to be fainter, with a significant fraction of the optical light coming from the host galaxy, thus improving the match to the galaxy templates. Uncertainties due to the mismatch between the observed AGN SEDs and the galaxy templates may be partially reflected in the redshift probability distributions, although the lack of a representative template may result in catastrophic failures, and an unrepresentative $p(z)$. Additionally, the priors are based on observations of the galaxy population in the HDF-N, and thus may not truly represent the AGN population. 

In Figure \ref{fig:photz} we compare the available spectroscopic redshifts with the best photo-$z$ estimates, given by the mode of the $p(z)$ distribution. Error bars on the photo-$z$ represent 95\% central confidence intervals determined from the $p(z)$ distribution, although will not fully represent more complicated multi-modal distributions. The estimates generally agree well at $z_\mathrm{spec}<1.5$, and have a $\delta z = |(z_\mathrm{phot}-z_\mathrm{spec})| / (1+z_\mathrm{spec}) \approx 0.18$. There are however a number of outliers. A high fraction of these are associated with broad-line QSOs, in which the optical light is dominated by the AGN, and thus cannot be accurately described with the BPZ galaxy templates. Excluding these sources reduces the error to $\delta z \approx 0.13$. The remaining outliers are generally at high redshifts ($z_\mathrm{spec}\gtrsim 1.2$), and are assigned large errors, indicating broad $p(z)$ distributions.
However, we note a systematic bias that assigns $z_\mathrm{spec}\gtrsim1.2$ sources to lower redshifts. The $p(z)$ distributions are equally affected by such bias; accounting for the redshift uncertainty, however broad, will not correct for this effect. Indeed, the systematic bias indicates a catastrophic failure of the BPZ code to correctly fit the observed SEDs with realistic templates at the higher redshifts. This failure may severely bias any determination of the XLF. We explore this in more detail in section \ref{sec:highz}.
%
\begin{figure}
\begin{center}
\includegraphics[width=\columnwidth]{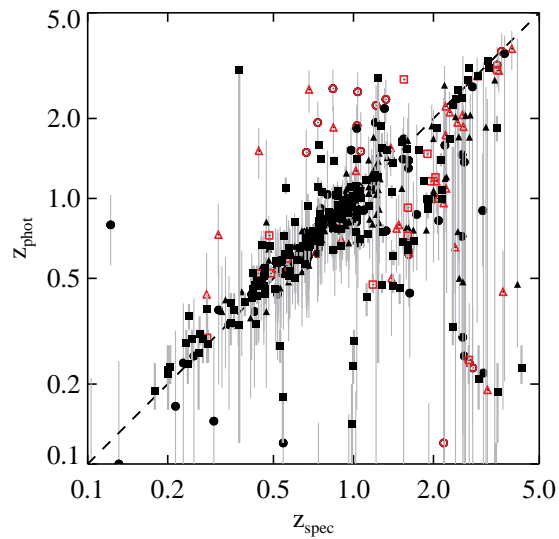}
\end{center}
\caption[Comparison of spectroscopic and photometric redshifts in the CDFs]{
Comparison of spectroscopic and template fitting photometric redshift estimates in the CDF-S (\textit{circles}), CDF-N (\textit{triangles}) and AEGIS (\textit{squares}). Error bars are 95\% central confidence intervals derived from the redshift probability distributions. At $z\lesssim1.0$ the photometric and spectroscopic redshifts are generally in good agreement. At higher redshifts the photo-$z$'s are less reliable, reflected by the larger error bars. A high fraction of the outliers are associated with broad-line QSOs (\textit{open symbols}), which are poorly fit by the templates. However, there is a systematic bias to assign $z\gtrsim 1.2$ sources to lower redshifts.
}
\label{fig:photz}
\end{figure}

\subsubsection{\chandra\ Deep Field-North}
\label{sec:cdfn}

The \chandra\ Deep Field-North (CDF-N) provides a second 2 Ms X-ray field. A point source catalogue based on the full 2 Ms exposure obtained has been presented by \citet{Alexander03}, however we use our own reduction and source catalogue, again using the procedure described in \citet{Laird09}. Our catalogue probes to lower significances than the \citet{Alexander03} work, and allows us to adopt our improved sensitivity calculations \citep{Georgakakis08}. In this paper we use objects with significant detections (false Poisson probability $<4\times 10^{-6}$) in the hard band, providing a sample of 303 X-ray sources.

A vast quantity of multiwavelength data are available in this thoroughly observed field \citep[e.g.][]{Dickinson03}. To find optical counterparts to our X-ray sources we use the Hawaii Hubble Deep Field North data \citep{Capak04}, which fully covers the X-ray data with deep observations in $U$, $B$, $V$, $R$, $I$, $z'$ and $HK'$ bandpasses. Optical counterparts were found using the likelihood ratio method, as in the CDF-S. This yields secure optical counterparts for 204 ($\approx 67$\%) of our hard-band selected X-ray sources.

We searched for spectroscopic redshifts for our secure optical counterparts in a variety of spectroscopic surveys that have been carried out in the region \citep{Barger08,Trouille08,Weiner06,Steidel03,Reddy06}. 
This provided spectroscopic redshifts 166 sources (\amend{81\% of the X-ray sources with secure optical counterparts}, 55\% of the \amend{total} hard X-ray selected sample). \amend{The spectroscopic completeness is much higher ($\sim 80$\%) for the brightest X-ray sources \citep[$\fx>10^{-14}$ \ergscm, see][]{Trouille08} }.

As for the CDF-S, we have recalculated photometric redshift estimates with BPZ, using the \citet{Capak04} $U$, $B$, $V$, $R$, $I$, $z'$ and $HK'$ photometry. A comparison of photometric and spectroscopic redshifts is shown in Figure \ref{fig:photz}. Again, a large number of the outliers are associated with broad-line AGN, in addition to high redshift sources which are assigned large errors. Excluding QSOs, the accuracy of our photo-$z$'s is $\delta z \approx 0.11$. A systematic bias at $z_\mathrm{spec}\gtrsim1.2$, assigning sources to lower redshifts, is present, as found for the CDF-S. 

\subsubsection{AEGIS}
\label{sec:aegis}

AEGIS \citep{Davis07} is an international multi-wavelength survey of the Extended Groth Strip, covering a $\sim$0.25 by 2 degree strip of the sky. Deep ($\sim 200$ ks) X-ray data have been obtained along the entire length of the strip. \citet{Laird09} described the reduction of these data, and presented a catalogue of 1325 unique point sources, merged over all bands. In this paper we restrict our analysis to hard band detected sources (false Poisson probability $<4\times 10^{-6}$) and those which also fall within the the area with very deep $u^*g'r'i'z$ optical data from the Canada-France Hawaii Telescope Legacy Survey\footnote{http://www.cfht.hawaii.edu/Science/CFHTLS}. This reduces the sample to 397 X-ray sources, of which 343 (86\%) have secure $i'$ band optical counterparts (likelihood ratio $>0.5$). 

Significant spectroscopic follow-up is available in the AEGIS field, including redshifts from the DEEP2 survey \citep{Davis07} and additional targeted follow-up of X-ray sources with \textit{MMT} \citep{Coil09}. This provides spectroscopic redshifts for 157 sources (40\%). 

In this field we have chosen to use photometric redshifts from 2 different sources. Huang et al. (in preparation) use an Artificial Neural Network/training set approach \citep[ANN$z$,][]{Collister04}. This approach effectively fits an arbitrary function to the redshift-colour relation, using a set of spectroscopic redshifts to train the neural network. As such the redshifts are only expected to be valid for sources with the same range of properties as the training set. 
The Huang et al. work uses an 3.6$\mu$m IRAC selected catalogue \citep{Barmby08}, further subjected to a number of colour cuts, and utilises photometry from \textit{MMT} (Megacam) as well as 
IRAC, MIPS, and DEEP2. The neural network was then trained using the DEEP2 spectroscopic data, and photometric redshifts estimated for all sources which satisfy the selection criteria. This provides highly accurate photo-$z$'s ($\delta z\approx 0.04$), albeit for a restricted sample. ANN$z$ redshifts are adopted for 38 X-ray sources in our sample (without spectroscopic redshifts). To account for the error we must determine a probability distribution for each redshift. For the ANN$z$ estimates we have based this on the distribution of spectroscopic redshifts, for a given ANN$z$ estimate, using all objects in the Huang et al. work with spectroscopic identification, as demonstrated in Figure \ref{fig:annz}.

\begin{figure}
\begin{center}
\includegraphics[width=\columnwidth]{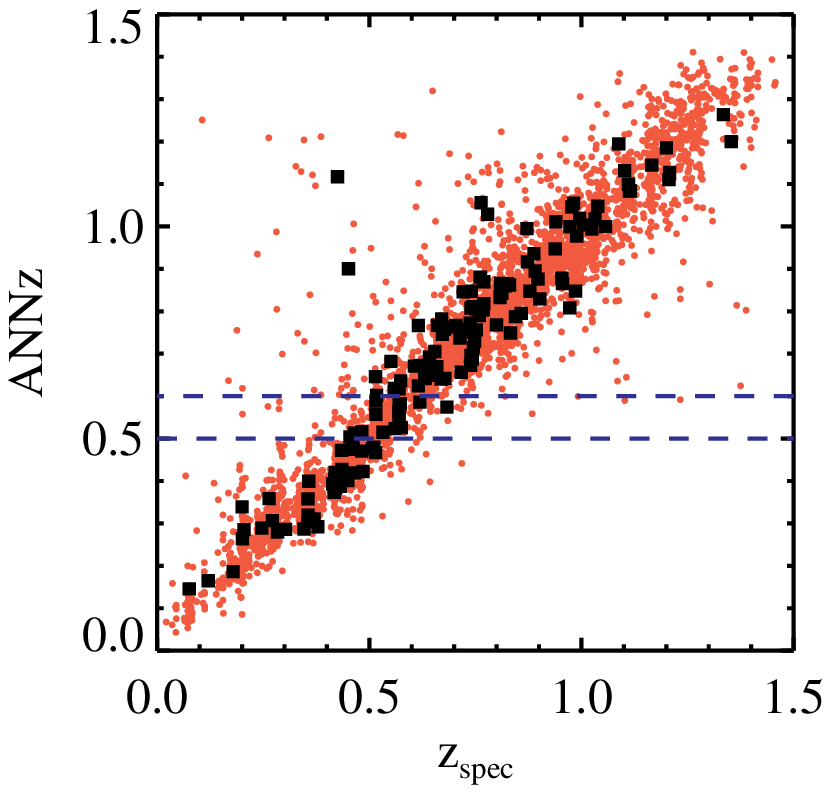}
\includegraphics[width=\columnwidth]{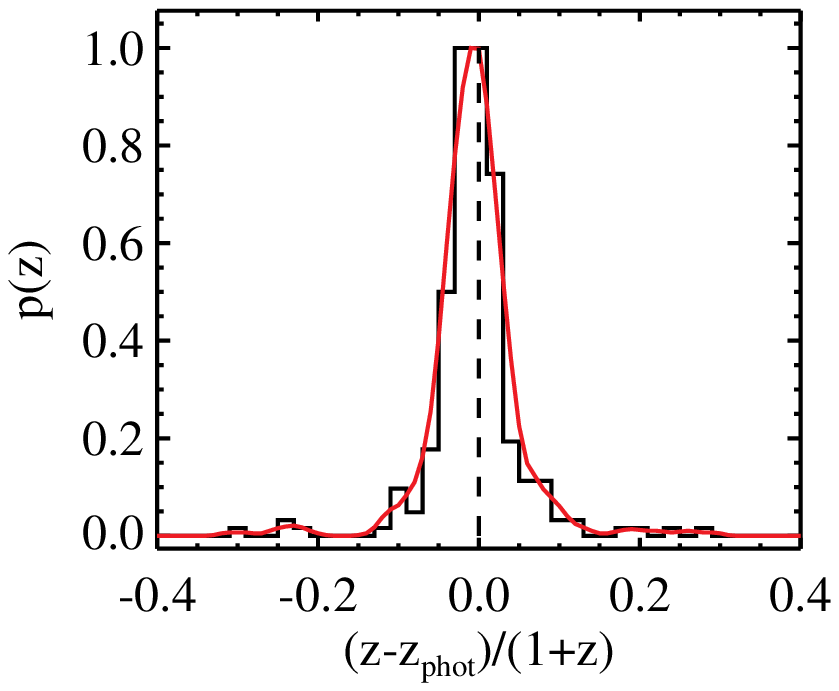}
\end{center}
\caption{
(\textit{Top}): Comparison of ANN$z$ and spectroscopic redshifts in AEGIS, shown for all sources (\textit{red circles}) and X-ray sources (\textit{black squares}). The ANN$z$ redshifts generally provide extremely good estimates, although are available for a limited sub-set of objects, and are restricted to $z<1.5$ where there are sufficient spectroscopic training data. The probability distribution for an ANN$z$ estimate is determined from the distribution of spectroscopic redshifts for a small slice of photometric redshifts (e.g. \textit{blue dashed lines}). (\textit{Bottom}): The distribution of $(z-z_\mathrm{phot})/(1+z)$ for the ANN$z$ redshift range indicated in the top figure. The true distribution (\textit{black histogram}) is smoothed (\textit{red curve}) to provide an estimate of probability distribution of true redshifts associated with a given ANN$z$ estimate.
}
\label{fig:annz}
\end{figure}


For the remaining sources in our sample with secure optical counterparts we adopt photometric redshifts from \citet{Ilbert06}, which are based on the CFHTLS $u^*g'r'i'z$ data. The redshifts are based on template fitting and a Bayesian approach, \amend{and thus use the same method as the BPZ code}. The priors are calibrated using spectroscopic identifications of galaxies in the D1 field of the CFHTLS. The estimates are less reliable than the ANN$z$ results ($\delta z\approx 0.16$), but do provide redshift information for the remaining optically identified X-ray sources. However there is a systematic bias for sources with $z_\mathrm{spec}\gtrsim1.2$ to be placed at lower redshifts, as was found for the CDF-S and CDF-N. The authors have made available the full $p(z)$ redshift probability distributions for every source, which we have adopted to account for the uncertainties, though once again will not account for the systematic bias.

\subsubsection{\textit{ASCA} Large Sky Survey}
\label{sec:alss}

The \textit{ASCA} Large Sky Survey \citep[ALSS,][]{Ueda99} covers a contiguous area of $\sim 5$ deg$^2$ near the north Galactic pole, and contains 34 sources detected in the hard (2--7 keV) band with the SIS detector, down to a flux limit $\sim 10^{-13}$ \ergscm\ (2--10 keV). 
Optical identifications have been presented by \citet{Akiyama00}. 2 of these sources are optically identified as galaxy clusters, 1 is a star and 30 are associated with AGN, all of which have spectroscopic redshifts. 1 source remains unidentified. An area curve for this survey was presented by \citet{Akiyama03}, which we adopt for our work. We note that this area curve does not fully account for the Poissonian nature of the X-ray source detection, but as a relatively high flux limit was set for the catalogue and the X-ray fluxes will be accurately measured this is unlikely to have an adverse affect on our bright end determination.

\subsubsection{\textit{ASCA} Medium Sensitivity Survey}
\label{sec:amss}

The largest area survey in our study is the \textit{ASCA} Medium Sensitivity Survey \citep[AMSS,][]{Ueda01}. We include sources from the AMSSn sub-sample, selected in the hard (2--10 keV) band, with optical identifications presented by \citet{Akiyama03}. The sample includes 87 X-ray sources, 76 of which are AGN, to a flux limit of $\sim 3 \times 10^{-13}$ \ergscm, and is 100\% optically identified with spectroscopic redshifts available for all the AGN. We include additional sources with redshift identifications from the AMSSs sub-sample (Ueda \& Akiyama, private communication), which includes 20 AGN; 2 sources in this sample remain unidentified. The area curve for the combined AMSSn and AMSSs samples was presented by \citet{Ueda03}.

\subsection{High-redshift sample}
\label{sec:highzdata}

Due to the catastrophic failure of our photometric redshift estimates at $z\gtrsim 1.2$, we have also investigated an alternative colour pre-selection approach for identifying AGN at high-redshift, building on the work of \citet{Aird08}. We include four of the fields used in \citet{Aird08}: \hdfn\ (fully encompassed by the CDF-N X-ray field described above), \lala, \lynx\ and \egs. We additionally include the large area covered by both the AEGIS-X survey and the CFHTLS field D3 optical imaging. This fully encompasses the \gws\ field used in \citet{Aird08}, and thus supersedes those data. Our high-redshift data overlaps with our hard X-ray selected samples described in section \ref{sec:hardxdata}; in section \ref{sec:highz} we explain how data from our hard X-ray surveys at $z<1.2$ and our high-redshift colour pre-selected samples are combined to determine the evolution of the XLF.

\subsubsection{Optical data}

The reduction of the optical data for \hdfn, \lala, \lynx\ and \egs\ and the source detection and photometry with SExtractor has been described in \citet{Aird08} and \citet{Reddy06}, and we refer the reader to those papers for further details.
For the CFHTLS we have used the publicly released images and catalogues available from Canada Astronomical Data Center\footnote{http://www4.cadc-ccda.hia-iha.nrc-cnrc.gc.ca/community/CFHTLS-SG/docs/cfhtls.html}, March 2008 release. The catalogues were generated using SExtractor, although with a slightly different configuration to \citet{Aird08}. Initial detection was performed in the $i'$-band (which is the deepest band), and magnitudes were determined in the other bands using the \citet{Kron80} style aperture (determined from the $i'$ band data). The data were not smoothed to match the seeing in different bands. These differences will affect the detection of objects and measurements of their colours, although given the depth and quality of the CFHTLS data is unlikely to adversely affect the colour selection of candidates. Nevertheless, any errors, biases and scatter due to the source detection and photometry procedure will be accounted for by our simulations to derive the optical selection functions (section \ref{sec:optsel}).

Samples of \z3 Lyman-Break Galaxies (LBGs) were extracted from the optical catalogues using the standard selection criteria from \citet{Steidel03}:
\begin{eqnarray}
\mathrm{G}-\mathcal{R} \leq 1.2 \nonumber\\
U_n-\mathrm{G} \geq \mathrm{G}-\mathcal{R}+1.0\nonumber\\
19.0\leq \mathcal{R} \leq 25.5
\label{eq:lbg}
\end{eqnarray}
with the faint limit of \R$\leq 25.0$ for \egs. An additional sample of objects at slightly lower redshifts is selected using the BX selection criteria \citep{Adelberger04}),
\begin{eqnarray}
\mathrm{G}-\mathcal{R} &\geq& -0.2 \nonumber\\
U_n -\mathrm{G} &\geq& \mathrm{G} -\mathcal{R} + 0.2 \nonumber\\
\mathrm{G} -\mathcal{R} &\leq& 0.2(U_n-\mathrm{G})+0.4\nonumber\\
U_n-\mathrm{G} &<& \mathrm{G}-\mathcal{R}+1.0
\label{eq:BX}
\end{eqnarray}
These objects are ``sub-U dropouts", falling in a thin slice of colour space below the Lyman-break criteria, with a selection function that peaks around \z2.5 (see section \ref{sec:optsel}). 

Objects were selected from the CFHTLS optical data using the same criteria. However, there are significant differences between the CFHTLS filter set and our \Un G\R\ filters used in the other fields. The \R-band falls between the $r'$ and $i'$ filters, thus total \R-band magnitudes were calculated for every object using an approximate colour correction, given by 
\begin{equation}
\mathcal{R}=r'+0.05-1.08(r'-i')
\end{equation}
which was determined from our model colour distributions described in section \ref{sec:optsel}.
These estimated \R-magnitudes were used to apply the total magnitude limits, consistent with our other fields. However, for the selection in colour-space we used the original $u^*g'r'$ magnitudes in place of \Un G\R, thus avoiding the effects of uncertain colour-corrections. The differences in the filters will lead to different selection functions, which we determine in section \ref{sec:optsel}.


\begin{table*}
\caption[List of fields with the numbers of colour-selected objects]{
Summary of the fields used to determine the high-redshift XLF using rest-frame UV colour pre-selection. }
\begin{center}
\resizebox{\textwidth}{!} {
\begin{tabular}{c c c c c c c c c c c c c c}
\hline
Field 		& RA      & Dec	     &  $\mathrm{N_H}$$^{a}$	 & X-ray 	       & Survey 		& Selection & $N_\mathrm{cand}$$^{d}$& $N_\mathrm{X}$$^{e}$ & $N_\mathrm{spec}$$^{f}$ & $N_{\mathrm{low}z}$$^{g}$ & $N_\mathrm{samp}$$^{h}$\\
name		&		&		     &	 						& exposure$^{b}$  	&   area$^{c}$  		& 	  	  	&		&		&            & $z<1.4$&		\\
			& (J2000) & (J2000) & ($10^{20} \mathrm{cm}^{-2}$)	    & (ks) 		& (arcmin$^2$) 	&       	 	& 		& 		& 		&           & 	\\	
\hline

HDF-N		& 12:36:55 & +62:14:18		&           1.5 				& 1862.9	& 149.1			&   BX           &   1192	& 	9	& 	8	&	2	&   7  \\
			&			  &				&		                            &                   &                         &    LBG        &   292   &     3     &     2     &    0      &  3   \\

AEGIS-X/CFHTLS & 14:17:43 &  +52:28:25 	 	&	1.2                       &   188.7       &  1155.6            &   BX           &  17050 &     77    &    26    &    10    &   66 \\
			&			  &				&		                            &                   &                         &    LBG        &   6933  &     30   &     10    &    0      &  30   \\

Lynx			& 08:48:56 &  +44:54:50 				&	1.0                       &  186.5        & 243.4                &   BX           &   647   &   2       &     0     &   0       &  2 \\
			&			  &				&		                            &                   &                         &    LBG        &   223   &    1      &     1      &   0       &  1  \\

EGS1		& 14:22:43 & +53:25:25       		  &     1.2                      &   177.8       & 358.9                &    BX          &   809   &     8    &     0       &   0      &  7  \\
			&			  &				&		                            &                   &                         &    LBG        &   329   &     7    &     0       &    0     &  7   \\

LALA CETUS& 02:04:44  &  $-$05:05:34 			&      2.2                      &   173.1       & 233.3                &   BX           &    660  &   4       &     0     &   0        &  4  \\
			&			  &				&		                            &                   &                         &    LBG        &    144   &   3       &     0     &   0       &   3  \\
\hline
\end{tabular}
}
\end{center}
\begin{flushleft}
$^{a}$Galactic column density from \citet{Dickey90}.\\
$^{b}$X-ray exposure time after good time interval and background ßare Þltering (average of the individual pointings for AEGIS-X).\\
$^{c}$Total survey area covered by X-ray and optical data.\\
$^{d}$Number of candidates from the optical imaging satisfying the colour selection criteria.\\
$^{e}$Number of the colour-selected candidates with secure X-ray counterparts.\\
$^{f}$Number of the X-ray detected candidates with spectroscopic redshifts.\\
$^{g}$Number of the spectroscopically identified sample which are low redshift interlopers ($z<1.4$).\\
$^{h}$Final number of sources in our sample after excluding spectroscopically-identified low-redshift interlopers and additional suspected low-redshift interlopers according to their position in the $\fx-f_\mathrm{opt}$ plane (see section \ref{sec:lowzinterlopers}, Figure \ref{fig:fxfopt_highz}).
\end{flushleft}
\label{tab:highzfields}
\end{table*}

\subsubsection{X-ray data}
\label{sec:xdat_highz}

All of our fields have been observed with \chandra, and have exposure times $\gtrsim 200$ ks. The data have been reduced using our pipeline procedure \citep{Laird09}. 
For our high-redshift samples we use sources detected in the soft-band with significance $<4\times 10^{-6}$. Our colour pre-selection limits us to sources at $z\gtrsim 2$, and at these redshifts the soft-band will probe rest-frame energies $\sim 2-8$ keV. Thus our selection should be relatively unbiased for the selection of moderately absorbed AGN ($\mathrm{N_H}\lesssim 10^{23}$ cm$^{-2}$), and is comparable with the hard-band selection employed at lower redshifts \citep{Cowie03,Barger05}.

\subsubsection{Matching the X-ray and optical samples}
\label{sec:xopt_correlate}

We determined which of our colour pre-selected high-redshift objects host AGN by matching the optical and X-ray source lists. 
The matching is performed with the \emph{entire} \R- (or $i'$-) selected optical catalogues using the likelihood ratio method. The colour selection criteria are then applied to the matches. Matching the entire optical catalogue ensures that X-ray sources are matched to the most likely optical counterparts, and prevents colour pre-selected objects being erroneously associated with nearby bright X-ray sources, which have correspondingly bright, secure optical counterparts \citep[cf. the direct extraction method used in ][]{Aird08}. 

In table \ref{tab:highzfields} we list the total numbers of X-ray detected objects within each of our colour-selected samples. We searched for spectroscopic redshifts for our sources as described in sections \ref{sec:cdfn} and \ref{sec:aegis} for the HDF-N and AEGIS-X fields respectively, and from \citet{Stern02} for the \lynx\ field. Remaining unidentified sources were assigned redshift probability distributions, $p(z)$, based on the optical selection functions (see section \ref{sec:optsel}).

\subsubsection{Low-redshift interlopers}
\label{sec:lowzinterlopers}

\Un G\R\ LBG selection at \z3 is known to be extremely clean and contains very few low redshift interlopers \citep{Steidel03}. The majority of interlopers are Galactic stars, which are not expected to be bright X-ray sources and thus will not contaminate our sample. However, the BX selection, which does not rely on very strong features such as the Lyman-break to identify candidates, is more susceptible to contamination from low-redshift sources, mainly from low-redshift ($z\lesssim0.3$) star-forming galaxies in which the Balmer-break is confused with the Lyman-$\alpha$ forest absorption at $z\gtrsim 2$ \citep{Adelberger04,Reddy06}. Indeed, a significant number of low-redshift ($z<1.2$) interlopers are identified within our spectroscopically observed, X-ray detected BX sample (see table \ref{tab:highzfields}, Figure \ref{fig:redshift_hist}). By examining the positions of these sources in the $\fx-f_\mathrm{opt}$ plane (see Figure \ref{fig:fxfopt_highz}) it is clear that 5 of these low-redshift interlopers occupy a distinct position, associated with optically-bright objects (\R$< 23$), but with relatively low X-ray fluxes, thus $\log (\fx/f_\mathrm{opt})<-1.5$. These objects are the expected bright, low-redshift star-forming galaxies, with X-ray emission associated with star-formation processes. We exclude 2 more X-ray detected BX sources from our analysis which fall in the same region and are thus also suspected to be low-redshift contaminants. The faint BX source with \R$\approx 23.5$ and $\log (\fx/f_\mathrm{opt})<-1.5$, which has a spectroscopic redshift in the correct range, is also excluded as we suspect a star-burst origin \citep[see also][]{Aird08}.
\begin{figure}
\begin{center}
\includegraphics[width=\columnwidth]{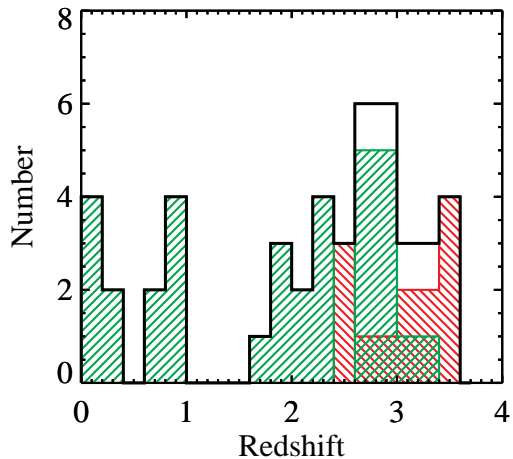}
\end{center}
\caption[Distribution of available spectroscopic redshifts for the X-ray detected BX and LBG samples]{
Distribution of available spectroscopic redshifts for the X-ray detected colour-selected objects. \textit{Green}: BX;  \textit{red}: LBG; \textit{black}: total. The BX sample is contaminated by low-redshift interlopers, although around half may be identified as low-redshift star-forming galaxies and on the basis of their $\fx/f_\mathrm{opt}$ ratio (see Figure \ref{fig:fxfopt_highz}, section \ref{sec:lowzinterlopers}). 
}
\label{fig:redshift_hist}
\end{figure}

\begin{figure}
\includegraphics[width=\columnwidth]{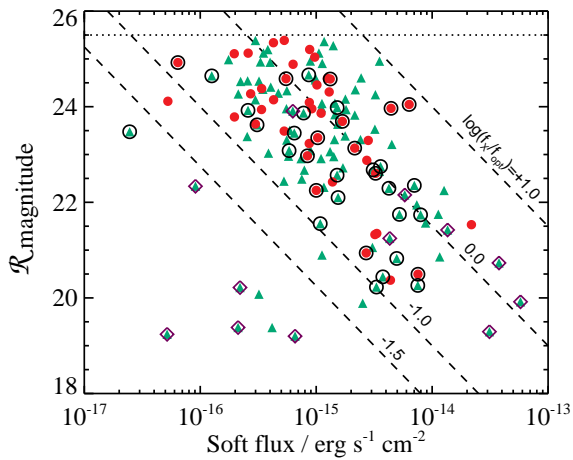}
\caption[Soft X-ray flux vs. \R-magnitude for our colour-selected samples]{
X-ray flux (0.5--2 keV) vs. \R-magnitude for the BX (\textit{green triangles}) and LBG (\textit{red circles}) X-ray detected samples. \textit{Black circles} indicate objects with spectroscopic redshifts, $z_\mathrm{spec}>1.4$; \textit{purple diamonds} indicate low-redshift interlopers with $z_\mathrm{spec}<1.4$. As discussed in section \ref{sec:lowzinterlopers} a number of the interlopers are associated with low-redshift star-forming galaxies, which are optically bright but very faint in X-rays. We thus exclude \emph{all} sources with $\log (\fx/f_\mathrm{opt}) < -1.5$. This will exclude the low-redshift interlopers, in addition to bright high-redshift star-burst galaxies that may also contaminate the AGN XLF at the faintest fluxes \citep[see ][]{Aird08}. Objects with $z_\mathrm{spec}<1.4$ are also excluded from our analysis, but interlopers associated with low-$z$ AGN may still contaminate the remaining spectroscopically unidentified BX-selected sample. 
}
\label{fig:fxfopt_highz}
\end{figure}

The 6 remaining interlopers, however, are associated with low-redshift AGN, and thus fall within the AGN region $-1.0\lesssim \log (\fx/f_\mathrm{opt}) \lesssim +1.0$. While any AGN which have been spectroscopically identified as low-redshift interlopers are excluded from our analysis, additional contaminants may remain within our unidentified BX sample. Based on our spectroscopic sample the fraction of low-redshift AGN contaminants may be as high as $\sim 25$\%, although our sample is biased in the sense that a higher fraction of bright objects have spectroscopic identifications. Indeed, in the large sample of BX-selected galaxies with spectroscopic redshifts used by \citet{Reddy08} the low redshift contamination in the \emph{galaxy} sample was $\sim 59$\% for bright candidates (\R$<23.5$), whereas for fainter candidates was only $\sim 9$\%. Conversely, in the sample of objects \emph{spectroscopically} identified as AGN (including QSOs and narrow-line AGN), the contamination fraction was only $\sim 12$\%, and approximately consistent over the entire magnitude range. Clearly a much larger, near complete sample of spectroscopic redshifts for BX/X-ray selected AGN is required to truly determine the interloper fraction within our sample, and could also allow improvements to the selection criteria (such as introducing data from redder wavebands) to be developed to help reduce the contamination. However, given the current uncertainties, we have chosen not to make further corrections to account for the interloper fraction within our sample. We thus note that the faint end of our XLF based on the BX sample may be over estimated due to low-redshift contamination.

\subsubsection{High luminosity sample}
\label{sec:highlx_highz}

The total area coverage for our high-redshift sample remains relatively small, and thus we require an additional sample of high-luminosity AGN to constrain the bright end of the XLF. We have chosen to use bright objects ($\fx> 5 \times 10^{-15}$ \ergscm) from the \citet{Hasinger05} compilation of surveys (excluding the CDF-N to avoid overlap with our own analysis). These samples provide a soft-band selected, highly complete, spectroscopically selected sample. We have included objects with $1.9<z\leq3.5$, providing 30 high-luminosity AGN in our high-redshift sample. These bright, complete, spectroscopic surveys may be combined with our colour pre-selected samples by assuming a top-hat selection function over the chosen redshift range (see section \ref{sec:evol_colsel}).

\section{A Bayesian approach to determine the XLF}
\label{sec:bayes_xlf}

Our compilation of X-ray surveys provides large samples of AGN over a range of redshifts with well-defined completeness, required to determine the evolution of the XLF. However, the faint sources from the \chandra\ fields are subject to uncertainty in both the redshift determination (for those sources with photometric redshifts or in the high-$z$ colour preselected samples) and the X-ray flux measurement (due to the small numbers of counts from the faintest sources). This has led us to develop a Bayesian methodology to make accurate inferences about the form and evolution of the XLF, accounting for the uncertainties in individual sources.

\subsection{The likelihood function}

The first step is to determine the likelihood function, $\mathcal{L}$, for which we have followed the work of \citet{Loredo04}. Our individual sources are effectively Poissonian points, drawn from the distribution given by the XLF. Thus, for a given model XLF, the expected number of detectable sources, $\lambda$, is given by 
\begin{equation}
\lambda = \int \dd \log L_\mathrm{X} \int \frac{\dd V}{\dd z} \dd z \; A(L_\mathrm{X}, z) \; \phi(L_\mathrm{X},z\; |\; \mathbf{\theta})
\label{eq:exnum}
\end{equation}
where the integrals are taken over the entire range of possible values. $A(L_\mathrm{X},z)$ is the area of the survey sensitive to a source of a particular luminosity, $L_\mathrm{X}$, and redshift, $z$, and thus a particular flux if we assume a single spectral shape (power law with photon index $\Gamma=1.9$).  $A(L_\mathrm{X},z)$ effectively describes the probability of a source being detected. The luminosity function, $\phi(L_\mathrm{X},z \;|\; \mathbf{\theta})$, is the differential number density of sources per unit co-moving volume as a function of $L_\mathrm{X}$ and $z$, given an assumed model described by the set of parameters, $\mathbf{\theta}$. $\frac{\dd V}{\dd z}$ is the differential co-moving volume per unit area, as a function of $z$.

As described in \citet{Loredo04}, the likelihood function may be constructed from the product of the probabilities of the observed data from the individual sources, given the model, and the probability that no other objects were detected. For a Poisson process the probability of no objects being detected is simply $e^{-\lambda}$. In the presence of individual source uncertainties, the probability of the observed data, $d_i$, from an individual source $i$, given the model XLF is found by marginalising $p(d_i \giv L_\mathrm{X}, z)$ over the model distribution of $\Lx$ and $z$. Thus,
\begin{eqnarray}
\mathcal{L}(\{d_i\} \;|\; \mathbf{\theta}) &=& \\
e^{-\lambda}\times \prod_{i=1}^M \int  &\dd \log L_\mathrm{X}& \int \frac{\dd V}{\dd z} \dd z \; p(d_i \;|\; L_\mathrm{X},z) \; \phi(L_\mathrm{X},z \;|\; \mathbf{\theta})\nonumber
\label{eq:xlflike} 
\end{eqnarray}
where $\{d_i\}$ represents the set of observed data from all sources, and the product in the second part of the equation is over $M$ detected sources. For the idealised situation where the redshift and luminosity are perfectly determined, $p(d_i \;|\; L_\mathrm{X},z)$ will be a delta function at the true values, and equation \ref{eq:xlflike} will generalise to the standard maximum likelihood estimator. For our sample $p(d_i \;|\; L_\mathrm{X},z)$ is the convolution of the redshift probability distribution $p(z)$ and the Poisson likelihood of obtaining the observed X-ray data, $N$ counts,
\begin{equation}
{\cal L}(N \giv s,b) = \frac{(s+b)^N}{N!}e^{-(s+b)}
\end{equation}
where $s$ is calculated from the hard band (2-10 keV) flux for a source of luminosity $L_\mathrm{X}$ and redshift $z$, assuming $\Gamma=1.9$ and Galactic $\mathrm{N_H}$. Thus the Poissonian nature of the counts from the X-ray sources, the associated uncertainty in the flux and the Eddington bias are naturally accounted for in our determination of the XLF. Assuming $\Gamma=1.9$ approximately corrects for absorption effects due to intrinsic column densities $\mathrm{N_H}\lesssim10^{23}$ cm$^{-2}$ at $z\sim1$, although we have not chosen to make a more sophisticated correction on a source by source basis (see further discussion in section \ref{sec:limitations}).
Our approach builds on the $\log N - \log S$ work of \citet{Georgakakis08}, extended to include redshift information, and allows us to use the improved method of calculating the X-ray sensitivity. For sources with spectroscopic redshifts we assume delta functions for $p(z)$. We also assume delta functions for the luminosities of sources in both the hard and soft X-ray selected, large area, bright samples (sections \ref{sec:alss}, \ref{sec:amss}, and \ref{sec:highlx_highz}).
\subsection{Correcting for incompleteness}
\label{sec:completecorr}

The likelihood function above does not account for the incompleteness of our samples. In our hard X-ray selected samples, between $\sim 14$\% (AEGIS) and $\sim 32$\% (CDF-N, CDF-S) of the sources lack secure optical counterparts, and thus have neither spectroscopic nor photometric redshifts. 
However, we can derive a completeness correction term to account for the incompleteness in the likelihood function. 

Figure \ref{fig:fxfopthard}(a) shows the distribution of the hard X-ray flux and optical magnitudes of our (optically identified) sources. Sources are scattered over a range of optical to X-ray flux ratios, approximately over the range $-1<\log f_\mathrm{X}/f_\mathrm{opt}<1$ although a small proportion are at lower ratios and a significant number of this hard X-ray selected sample exhibit $\log f_\mathrm{X}/f_\mathrm{opt}>1$. At faint X-ray fluxes, a large fraction will be unidentified due to their faint optical magnitudes. In Figure \ref{fig:fxfopthard}(b) we show the $\log f_\mathrm{X}/f_\mathrm{opt}$ distribution of our sources. A Gaussian approximation, taking the sample mean, $\mu$, and standard deviation, $\sigma$, is shown in red, and provides a reasonable description of this distribution. If we apply a strict magnitude limit of $R=25.5$ (sources fainter than this limit are now deemed unidentified), and assume that all the unidentified sources would have optical counterparts fainter than this limit (reasonable given the very deep optical data we have used to search for counterparts), then we can introduce a correction to the expected number of sources, $\lambda$, given by equation \ref{eq:exnum}, based on the fraction of the $\log f_\mathrm{X}/f_\mathrm{opt}$ distribution probed at a given flux. Thus,
\begin{eqnarray}
\lambda =\;\;\;\;\;\;\;\;\;\;\;\;\;\;\;\;\;\;\;\;\;\;\;\;\;\;\;\;\;\;\;\;\;\;\;\;\;\;\;\;\;\;\;\;\;\;\;\;\;\;\;\;\;\;\;\;\;\;\;\;\;\;\;\;\;\;\;\;\;\;\;\;\;\;\;\;\;\;\;\;\;\;\;\;\;\;\;\;&&\\
\int \dd \log L_\mathrm{X} \int {\textstyle \frac{\dd V}{\dd z} \dd z} \int \dd R\; A(L_\mathrm{X}, z) \phi(L_\mathrm{X},z | \mathbf{\theta})  g(\Lx,z,R | \mu,\sigma)&&\nonumber\\
= \int \dd \log L_\mathrm{X} \int {\textstyle \frac{\dd V}{\dd z} \dd z} \; A(L_\mathrm{X}, z) \; \phi(L_\mathrm{X},z\; |\; \mathbf{\theta}) \; C(\Lx,z \;|\; \mu,\sigma)&&\nonumber
\label{eq:corrlambda}
\end{eqnarray}
where $g(\Lx,z,R \giv \mu,\sigma)$ describes the $R$-magnitude distribution of sources at a given flux, $\fx(\Lx,z)$, and is derived from the Gaussian distribution of $\log \fx/f_\mathrm{opt}$. $g(\Lx,z,R \;|\; \mu,\sigma)$ is normalised such that
\begin{equation}
\int_{-\infty}^{\infty} \dd R \; g(\Lx,z,R \;|\; \mu, \sigma) = 1.
\nonumber
\end{equation}
Thus the factor 
\begin{equation}
C(\Lx,z \giv \mu,\sigma)=\int_{-\infty}^{25.5} \dd R \;g(\Lx,z,R \;|\; \mu,\sigma)
\end{equation}
is a completeness correction, reducing the expected number of detectable sources with low X-ray fluxes.

\begin{figure}
\begin{center}
\includegraphics[width=\columnwidth]{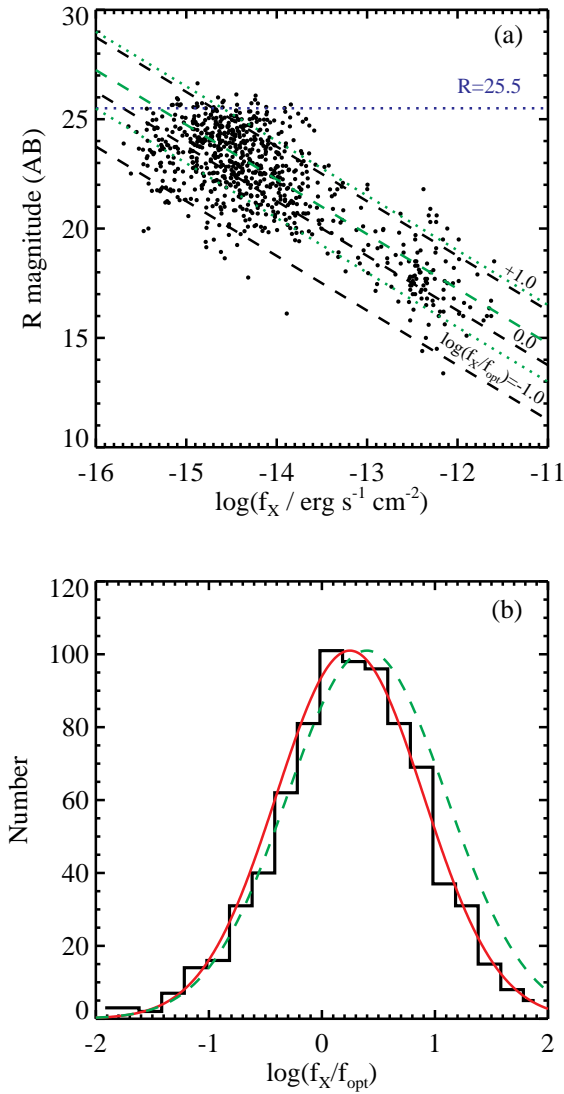}
\end{center}
\caption[The $f_\mathrm{X}$-$f_\mathrm{opt}$ distribution of our hard-band selected sample]{
(a) Hard (2--10 keV) flux against $R$ magnitude (AB) for our entire hard X-ray selected sample (\textit{black circles}). {\it Black dashed lines}: $\log \fx/f_\mathrm{opt}=$-1.0, 0.0 and +1.0, indicating the typical range of AGN. {\it Green lines:} mean (\textit{dashed}) and $\pm1\sigma$ ({\it dotted}) of the best-fit Gaussian $\log \fx/f_\mathrm{opt}$ distribution (see section \ref{sec:results}). {\it Blue dotted line}: $R=25.5$ limit for including optical counterparts. (b) Histogram of the distribution of $\log \fx/f_\mathrm{opt}$ ({\it black}), and the Gaussian distribution corresponding to the sample mean and standard deviation ({\it red curve}). This distribution will be biased due to incomplete optical identifications, as well as the distribution of X-ray fluxes. We account for these effects when fitting our evolutionary models (see sections \ref{sec:completecorr} and \ref{sec:xlf_evol}), and derive the $\log \fx/f_\mathrm{opt}$ distribution ({\it green dashed curve}), which is shifted to higher $\fx/f_\mathrm{opt}$ ratios.
}
\label{fig:fxfopthard}
\end{figure}

However, the observed distribution (the red curve in Figure \ref{fig:fxfopthard}) will be biased by the very incompleteness we are trying to correct for, as sources with $R\gtrsim 25.5$ will be missed. The extent of the bias will depend on the flux distribution of the X-ray sources (and thus the XLF, and the X-ray sensitivity), which determines the number of sources with faint X-ray fluxes where the problem arises. Therefore, to improve our completeness correction the $\log \fx/f_\mathrm{opt}$ distribution should be determined simultaneously with the XLF, accounting for the incompleteness, using the observed data (the $R$-magnitudes of the detected sources). We therefore introduce $g(\Lx,z,R)$ into the second term in equation \ref{eq:xlflike}, modifying the likelihood to
\begin{eqnarray}
\mathcal{L}(\{d_i\} \;|\; \mathbf{\theta}, \mu, \sigma) = e^{-\lambda} \times \;\;\;\;\;\;\;\;\;\;\;\;\;\;\;\;\;\;\;\;\;\;\;\;\;\;\;\;\;\;\;\;\;\;\;\;\;\;\;\;\;\;\;\;\;\;\;\;\;\;\;\; &&\\
\prod_{i=1}^M \int \dd \log \Lx \int {\textstyle \frac{\dd V}{\dd z}} \dd z \; p(d_i | L_\mathrm{X},z)  \phi(L_\mathrm{X},z | \mathbf{\theta})  g(\Lx,z,R_i | \mu,\sigma)&&\nonumber
\label{eq:corrlike} 
\end{eqnarray}
where we neglect any uncertainty in the measurement of the $R$-magnitude of an individual source, $R_i$, which is a minor simplification. The parameters $\mu$ and $\sigma$ are introduced as nuisance parameters, and will be determined along with the XLF, integrating over their uncertainties when making inferences regarding the XLF parameters, $\mathbf{\theta}$. In Figure \ref{fig:fxfopthard}b the green dashed line indicates the best-fit (mode) values of $\mu=0.24\pm0.01$ and $\sigma=0.68\pm0.01$ for our LADE evolutionary model (section \ref{sec:results}). The peak of the distribution is clearly shifted to a higher $\fx/f_\mathrm{opt}$, indicating the bias in the simple estimates.

\subsection{Incorporating the high-$z$ colour pre-selected samples}
\label{sec:highz}

In section \ref{sec:xlf_evol} we present results for the evolution of the XLF over the full available redshift range based on our hard X-ray selected sample only (using spectroscopic and photometric redshifts). However, due to the systematic bias in photo-$z$ estimates at $z\gtrsim 1.2$, we also wish to derive the evolution using our high-redshift colour pre-selected samples. Due to the a priori colour selection these samples will suffer from large, but well-defined incompleteness. Accounting for this incompleteness, and combining our soft-band selected high-$z$ sample with the hard X-ray selected samples at lower redshifts requires a number of further modifications to our methodology.  

\subsubsection{Optical selection functions}
\label{sec:optsel}

The first step is to determine the optical selection functions for the colour selection criteria. These were determined using the simulation method described in \citet{Aird08} \citep[see also][]{Steidel99, Hunt04, Reddy08}. Three sets of template spectra were determined for different spectroscopic classifications \citep[QSO, NLAGN or GAL: see ][]{Steidel03}, and used to generate model \Un G\R and $u^*g'r'$ colour distributions. GAL and NLAGN colours were generated using a \citet{Bruzual03} model template spectrum for a galaxy with continuous star formation and age 1 Gyr. This template spectrum was reddened using the \citet{Calzetti00} relation for obscuration by dust, with various extinction coefficients drawn from the distributions based on the observed range of UV spectral slopes for GAL and NLAGN sources in LBG samples \citep{Adelberger00,Steidel02,Steidel03}. Such modelling is not a full physical description for the NLAGN, being based only on a template spectrum for star-forming processes, but is sufficient to simulate the observed range of spectral shapes and redder distribution of broadband colours for NLAGN. For QSOs we have followed the work of \citet{Hunt04}. The template spectra were generated by varying the continuum slope and Ly$\alpha$ equivalent widths of a composite of 59 QSO spectra \amend{(corrected for intergalactic absorption)} from \citet{Sargent89}, based on the Gaussian distributions given in \citet{Hunt04}. 
\amend{These model spectra for all three types were then redshifted over the range $z=1-4$ and intergalactic absorption was applied by simulating a random line-of-sight to each object with intervening Lyman-limit absorption systems distributed according to the MC-NH model of \citet{Bershady99}, as described by \citet{Hunt04}. The final spectra are multiplied by the filter transmission curves to produce the model colour distributions.}
Figure \ref{fig:coldistcomp} compares colours for a single model spectrum (for each classification) at different redshifts in both \Un G\R\ and $u^*g'r'$ colour space. 

The full colour distributions were used to add artificial objects into the optical imaging data. Selection functions are then determined as the probability of recovering an object with a given \R\ magnitude with either the LBG or BX selection as a function of redshift. Objects were recovered using our SExtractor routines \citep[see ][]{Aird08} in the \Un G \R\ fields. The SExtractor configuration file used to generate the released catalogues for CFHTLS is available on-line\footnote{http://www1.cadc-ccda.hia-iha.nrc-cnrc.gc.ca/community/CFHTLS-SG/docs/cfhtlscat.sex}, thus we used this configuration for those data. Objects were selected in the $i'$-band image and flexible \citet{Kron80} style apertures were used to measure the magnitudes in other bands, and calculate the output colours. Our selection functions will thus account for the different source detection and photometry technique, and the effects of the different filter sets. 
\begin{figure*}
\includegraphics[width=\textwidth]{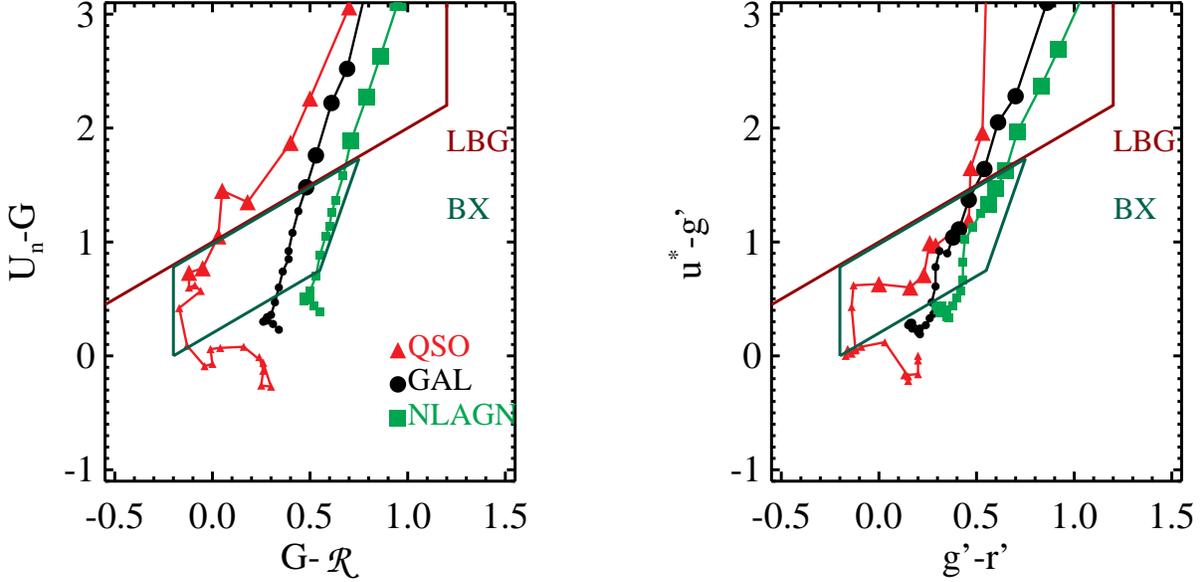}
\caption[Model colour distributions for our GAL, NLAGN and QSO classifications in \Un G\R, $u^*g'r'$ and $g'r'i'$ colour-space]{
Example model colour distributions for our GAL (\textit{black circles}), NLAGN (\textit{green squares}) and QSO (\textit{red triangles}) classifications.
The same model for each type is shown in 
\Un G\R\ (\textit{left}) and $u^*g'r'$ (\textit{right}) colour-space. The model is redshifted over $1.0<z<4.0$ Marks show the colours at $\Delta z=0.1$ intervals, with larger symbols indicating $2.5<z<3.5$. The variation between colours through different filter sets leads to differences in the optical selection functions (see Figure \ref{fig:highzselfunc}, section \ref{sec:optsel}).
}
\label{fig:coldistcomp}
\end{figure*}

Figure \ref{fig:highzselfunc} compares our final selection functions (derived from the simulations) for the different colour criteria, filter sets, and optical classifications. There are clear differences between the \Un G\R\ and $u^*g'r'$ selections, which we expect given the varying wavelength coverage of the filter sets, and results in variation in our selection functions. The $u^*$ filter extends to longer wavelengths than the \Un\ filter, and as such the $u^*-g'$ colour becomes sensitive to the Lyman-break at higher redshifts than the \Un$-$G colour, as can be seen from the tracks in Figure \ref{fig:coldistcomp}. Thus the LBG selection functions are shifted to slightly higher redshifts when using $u^*g'r'$ rather than the \Un G\R\ filter sets. The same effect leads to BX selection functions which probe to higher redshifts, although the low redshift cut-off is relatively unaffected. Thus the BX selection functions are generally broader for $u^*g'r'$ selection.

The red continuum of our NLAGN model generally leads to a reduced efficiency of the selection functions. This effect is particularly seen for the BX selection function for \Un G\R\ colours. The example NLAGN model shown in Figure \ref{fig:coldistcomp} has colours which are very close to the boundary of the BX selection box. The intrinsic scatter we apply to our models and the photometric scatter introduced via our simulations will both lead to incompleteness of the BX selection function, which is clearly seen in Figure \ref{fig:highzselfunc} (bottom centre panel). However, the $r'$ filter is significantly bluer than the \R\ bandpass, resulting in $g'-r'$ colours which are generally bluer than G$-$\R, and thus the NLAGN BX selection function is less adversely affected by the red continuum for $u^*g'r'$ selection, and has a higher efficiency (cf. top and bottom central panels in Figure \ref{fig:highzselfunc}). The same effect is seen for the GAL BX selection functions, although is much milder as the distribution of GAL colours is not as red as the distribution of NLAGN colours.

The model tracks for the QSO colours in Figure \ref{fig:coldistcomp} follow complicated routes which have very different shapes for the different filter sets. This is due the presence of the strong, broad emission lines, which will move in and out of the various filters at different redshifts. However, the basic characteristics necessary for our colour selections remain present: the Lyman-break leads to a rapid reddening of the \Un$-$G (or $u^*-g'$) colour at \z3; a characteristically flat spectrum is found at lower redshifts.

\begin{figure*}
\begin{center}
\includegraphics[width=\textwidth]{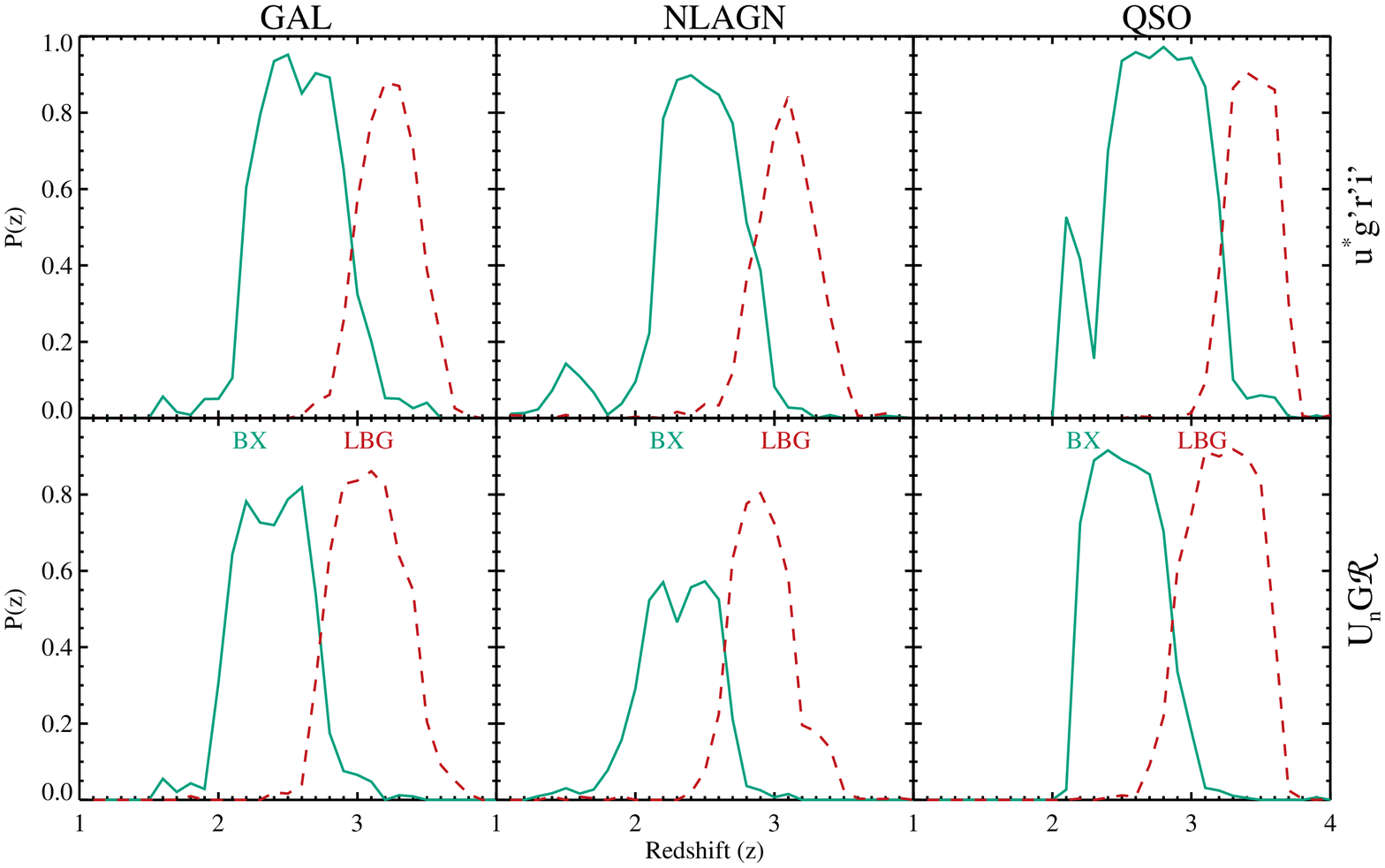}
\end{center}
\caption[Optical selection functions for our different colour criteria, optical classifications and filter sets.]{
Optical selection functions for our BX (\textit{green solid lines}) and LBG (\textit{red dashed lines}) colour selection criteria. The top row shows the selection functions for the CFHTLS-D3 optical data and $u^*g'r'i'$ filter sets. The bottom row shows selection functions in the \hdfn\ (with optical data of comparable depth) using \Un G\R\ filters. Each column is for a different spectroscopic classification as indicated. Significant differences in efficiency and redshift coverage exist between the selection functions for different optical types and filter sets (see discussion in section \ref{sec:optsel}).
}
\label{fig:highzselfunc}
\end{figure*}

\subsubsection{Modifying the likelihood function}
\label{sec:evol_colsel}

The optical selection functions described above can be used to modify the likelihood function derived in section \ref{sec:completecorr}, to allow us to combine the hard X-ray selected samples at low redshifts ($z<1.2$) and our colour pre-selected high-redshift samples. 
Firstly, $A(\Lx,z)$, which effectively gives the probability of detecting a source with given $\Lx$ and $z$, must be modified to account for the 2 different selection techniques. For $z<1.2$, $A(\Lx,z)$ is calculated from the area curves for our hard X-ray surveys, converting the hard (observed 2--10 keV) X-ray flux to an $\Lx$ (rest-frame 2-10 keV) and $z$ assuming $\Gamma=1.9$ and Galactic $\mathrm{N_H}$, as before. For $z>1.2$, we use the area as a function of soft X-ray flux, multiplied by the average of the combined BX and LBG selection functions, $\tilde{P}_f(z)$ for all magnitudes and spectroscopic classifications for a field, scaled to a maximum of 1. Thus,
\begin{eqnarray}
\tilde{P}_f(z) &\propto& \sum_\mathrm{types} \int_{19.0}^{25.5} P_f(\mathcal{R},z) \dd\mathcal{R} \\
A(\Lx,z)&=& \sum_\mathrm{fields} A_\mathrm{f}(\fx(\Lx,z)) \times \tilde{P}_f(z) \;\; [z>1.2]
\label{eq:highz_acurve}
\end{eqnarray}
where $A_\mathrm{f}(\fx(\Lx,z))$ is the area of each field sensitive to a soft (0.5--2 keV) flux for a given 2--10 keV luminosity source at $z$, assuming $\Gamma=1.9$ and Galactic $\mathrm{N_H}$, for each field. 
This reduces the probability of objects being detected outside of the range covered by the colour pre-selection, and thus provides realistic expected numbers at high redshifts in the likelihood evaluation (equation \ref{eq:corrlike}). As we consider the uncertainty in the X-ray flux we can utilise the method of \citet{Georgakakis08} to determine the X-ray area curves. For the high luminosity sample (section \ref{sec:highlx_highz}), the equivalent selection function is simply a top-hat function over the entire redshift range ($1.9<z<3.5$), corresponding to a complete spectroscopic follow-up of the purely X-ray selected high-luminosity sample. 

For each of the individual colour-selected sources we must assume a redshift probability distribution, $p_i(z)$ for which we adopt the selection function for the appropriate field and sample (unless a spectroscopic redshift is available in which case we adopt a delta function). 
Thus the differing redshift coverage of the selection criteria for different types and filter sets will provide some additional information regarding the evolution, and this effect will be explicitly accounted for by our study. The X-ray flux uncertainty is also accounted for in an identical manner as for the low-redshift hard-band selected samples, only converting from the soft-band flux to 2--10 keV rest-frame luminosities, to determine $p(d_i \giv \Lx,z)$ for each of our colour selected objects.

Additional weighting factors, $w_i$, must also be introduced, which are added as exponents of the individual source likelihoods in equation \ref{eq:corrlike}. This factor serves two purposes. Firstly, it reduces the weight assigned to sources in our hard X-ray samples according to the fraction of their $p(z)$ distribution which falls below $z<1.2$.
\begin{equation}
w_i=\frac{\int_0^{1.2} \dd z \; p_i(z)}{\int_0^\infty \dd z \; p_i(z)}.
\end{equation}
Thus, sources will be assigned a weight of $w_i=1$ if they have a spectroscopic redshift $z_\mathrm{spec}<1.2$, or if their $p_i(z)$ based on the photo-$z$ estimate falls wholly below $z=1.2$. Sources with $z_\mathrm{spec}>1.2$ or $p_i(z)$ which is all above $z=1.2$ will be assigned a weight of 0 (effectively removing them). Sources with photo-$z$ estimates that span above and below $z=1.2$ will be given appropriate fractional weights.

For the colour pre-selected objects, $w_i$ is the ratio of the total co-moving volume sampled based on our $z>1.2$ estimate of the area curve (equation \ref{eq:highz_acurve}), and the 
the effective co-moving volume sampled, based on the selection function at the object's \R\ magnitude. 
\begin{equation}
w_i =\frac{V_{\mathrm{tot}}} {V_{\mathrm{eff}}} =
	\frac{ \int_{1.2}^{\infty} A(L_i,z) \frac{\dd V}{\dd z} \; \dd z }
		{ \sum_{\mathrm{fields}}\int A_f(L_i,z) P_f(\mathcal{R}_i,z) \frac{\dd V}{\dd z} \dd z }
\label{eq:wi_new}
\end{equation}
where $P_f({\cal R}_i,z)$ is the selection function for both BX and LBG selection for a field for the observed \R-magnitude. Thus for the high-redshift sample, the $w_i$ factors apply additional weight to  sources with fainter optical magnitudes, which were thus less likely to be selected by the colour pre-selection technique. 

Our colour-selected sources are also incorporated into the calculation of the $\fx/f_\mathrm{opt}$ distribution, and the subsequent completeness correction. The derived $\fx/f_\mathrm{opt}$ considers hard-band X-ray fluxes. To achieve this an equivalent hard-band flux is calculated for each $\Lx,z$, whereas $\Lx$ for a source was originally derived from the observed soft-band flux. $R$-band magnitudes are derived from the \R-magnitudes by applying a small colour correction (based on our model templates described above).

Sources identified as low-redshift interlopers (section \ref{sec:lowzinterlopers}) are excluded from the colour-selected samples, but may still be included in the main sample to constrain the $z<1.2$ XLF for the AEGIS and CDF-N fields where our surveys overlap. 34 sources which were assigned photometric redshifts in our original hard X-ray selected samples are also present in our BX or LBG samples. Of these 17 were assigned $z_\mathrm{phot}<1.2$; the remainder were assigned to the higher redshifts expected for objects in our colour selected samples.
For all of these sources we do not consider the photo-$z$ distribution, and consider them solely with the colour selected samples, assigning the associated redshift distribution (based on the optical selection functions) and weighting factors. 


\subsection{Bayesian model comparison and parameter estimation}
\label{sec:nestsamp}

Having determined the likelihood function, our knowledge of the XLF can be fully described by the posterior probability function,
\begin{equation}
p(\theta, \mu, \sigma \;|\; \{d_i\},H) = \frac{\mathcal{L}(\{d_i\} \;|\; \mathbf{\theta}, \mu, \sigma) \; \pi(\mathbf{\theta}, \mu, \sigma \giv H)}
	{p(\{d_i\} \;|\; H)}
\end{equation}
where we have introduced the symbol $H$ to represent a particular model hypothesis for the XLF, and $\pi(\mathbf{\theta}, \mu, \sigma \giv H)$ is the prior probability we assign to values for the parameters $\theta$, for model $H$. The denominator, $p(\{d_i\} \;|\; H)$, is known as the Bayesian evidence, $\mathcal{Z}$, and normalises the posterior. 

There are now two tasks at hand. Firstly, we wish to determine the best estimates of the parameters, $\theta$, for our model XLF, and assign them an error.
Standard Monte-Carlo Markov Chain approaches can be used to draw samples from the posterior, and thus provide estimates of the parameters. Such approaches ignore the normalising factor, $p(\{d_i\} \;|\; H)$, as it is independent of the parameters. 
Our second task is to compare different evolutionary models. Bayesian model comparison is achieved by calculating the ratio of the posterior probabilities for different models \citep[e.g.][]{Kass95},
\begin{equation}
\frac{p(H_1 \;|\; \{d_i\})} {p(H_2 \;|\; \{d_i\}) } = \frac{ p(\{d_i\} \giv H_1) }{ p(\{d_i\} \giv H_2) } \; \frac{\pi(H_1)}{\pi(H_2)} 
	= \frac{\mathcal{Z}_1}{\mathcal{Z}_2} \;  \frac{\pi(H_1)}{\pi(H_2)} 
\end{equation}
where the ratio of the prior probabilities of the two different models, $\pi(H_1)/\pi(H_2)$, can be taken as unity if we have no prior knowledge which indicates we should favour one of the models. The evidence can be calculated by integrating the likelihood over the prior parameter space, thus,
\begin{equation}
\mathcal{Z}=\int \dd^D \theta \; \mathcal{L}(\{d_i\} \;|\; \mathbf{\theta}) \; \pi(\mathbf{\theta} \;|\; H)
\label{eq:evidence}
\end{equation}
where $D$ represents the number of dimensions for the integration (i.e. the number of parameters, now incorporating the nuisance parameters $\mu$ and $\sigma$ into the parameter set $\theta$). To perform this integration we have implemented the ``nested sampling" algorithm \citep{Skilling04}, which provides an estimate of the evidence, $\mathcal{Z}$, along with a numerical uncertainty. 800 ``live" objects are used for the nested sampling to ensure that the large, multi-dimensional parameter space is well sampled. Estimates of the posterior distribution may be recovered as a by-product of the nested sampling procedure, and thus used to determine the values of the parameters describing the XLF and their errors. We choose to report the posterior mode (equivalent to maximum likelihood) values of the parameters as our best estimates, and $1\sigma$ equivalent (68.3\%) Highest Posterior Density (HPD) intervals, which minimise the size of the confidence interval for the given confidence level. 

It is worth noting the difference of this Bayesian model comparison approach to frequentist methods. The frequentist approach to compare two models would be to calculate the likelihood ratio, the ratio of the maximum likelihood values for the two different models \citep{Neyman28}. 
The Bayesian approach, on the other hand, integrates the likelihood over the entire prior parameter space. As such the Bayesian evidence incorporates knowledge of all the uncertainties in the data and the parameter estimation. This approach also has the advantage that it naturally implements Occam's Razor, favouring a simple model with a compact parameter space, thus less free parameters, over a more complicated one, unless the latter provides a significantly better description of the data \citep{Berger92}.

\subsection{Binned estimates}
\label{sec:binned}

Our analysis of the evolution of the XLF considers each detected object individually, and does not require any binning. Our results consist of our best-fit parameter estimates and their uncertainties, and the relative Bayesian evidences for the different evolutionary models presented in section \ref{sec:xlf_evol} and tables \ref{tab:xlf_evol} and \ref{tab:evol_colsel}. However, for display purposes it is useful to produce binned estimates which represent the value and uncertainties over certain discreet ranges of luminosity and redshift. Our binned estimates are based on the $N_\mathrm{obs}/N_\mathrm{mdl}$ method \citep{Miyaji01}, which multiplies the value of the best-fit model evaluated at the centre of the bin by the ratio of the number of observed sources, $N_\mathrm{obs}$, to the predicted number in the bin, $N_\mathrm{mdl}$, where
\begin{equation}
N_\mathrm{mdl}= \int_{\Delta \log \Lx} \dd \log \Lx \int_{\Delta z} \frac{\dd V}{\dd z} A(\Lx,z) \phi(L_\mathrm{X},z | \mathbf{\theta}) 
\end{equation}
for each bin of size $\Delta \log \Lx$ and $\Delta z$.
With our methodology a source is assigned a distribution of luminosities and redshifts, and thus can be included in multiple bins. Therefore the observed number of sources, $N_\mathrm{obs}$, is found by summing the fractional contribution of all sources to each $\Delta \Lx-\Delta z$ bin,
\begin{equation}
N_\mathrm{obs}=\sum_{i=1}^M \frac{ \int_{\Delta \log \Lx} \dd \log \Lx \int_{\Delta z} \frac{\dd V}{\dd z} \; p( d_i \giv \Lx,z) }
								  { \int_{-\infty}^{\infty} \dd \log \Lx  \int_0^{\infty} \frac{\dd V}{\dd z} \; p( d_i \giv \Lx,z) }.
\end{equation}
Errors on the binned values are based on approximate Poisson uncertainties on the effective number of objects in a bin. Points are only shown for bins with an effective $N_\mathrm{obs} \geq 1$.

\section{Evolution of the XLF}
\label{sec:xlf_evol}

We have used our Bayesian methodology and the nested sampling algorithm to compare three different evolutionary models of the XLF, using both our hard X-ray selected samples over all redshifts (using photometric and spectroscopic redshifts), and the combination of the hard X-ray sample at $z<1.2$ and the high-redshift colour pre-selected samples. For each evolutionary model we calculated the best-fit (mode) for the parameters, with errors given by the Highest Posterior Density interval, and the relative logarithmic evidence, $\Delta \ln \mathcal{Z}$. Differences in $\ln \mathcal{Z}\gtrsim 4.6$ indicate very strong evidence in favour of the model with the higher evidence \citep{Jeffreys61}. The results are given in tables \ref{tab:xlf_evol} and \ref{tab:evol_colsel} for the hard X-ray sample only and the combined samples respectively. Constant (or log-constant) priors were assumed for all parameters over the ranges indicated in table \ref{tab:xlf_evol}. 

\subsection{Evolutionary models}
We compared the following evolutionary models:

\subsubsection{Pure Luminosity Evolution (PLE)}

At first we investigated models in which the XLF retains the same shape at all redshifts, but undergoes overall shifts in luminosity. We assume the standard double power-law shape for the XLF,
\begin{equation}
\phi(L_\mathrm{X})= \frac{\dd \Phi(L_\mathrm{X})}{\dd \log L_\mathrm{X}} = K_\mathrm{norm} \left[  \left(\frac{L_\mathrm{X}}{L_*} \right)^{\gamma_1} + \left(\frac{L_\mathrm{X}}{L_*}\right)^{\gamma_2} \right]^{-1}
\label{eq:lf}
\end{equation}
where $\gamma_1$ is the faint-end slope, $\gamma_2$ is the bright-end slope, $L_*$ is the characteristic break luminosity and $K_\mathrm{norm}$ is a normalisation factor for the overall density.
Pure luminosity evolution can then be described as evolution of the characteristic luminosity, $L_*(z)$. \citet{Barger05} found that the low redshift ($z<1.2$) XLF could be well described by a PLE model, where $L_*(z)$ is given by
\begin{equation}
\log L_*(z) = \log L_0 + p \log \left( \frac{1+z}{2} \right)
\label{eq:barPLE}
\end{equation}
However, previous studies \citep[e.g.][]{Ueda03,Silverman08,Aird08} have indicated that the strong evolution of $L_*$ at low redshifts may not continue above \z1. A number of authors have utilised models which extend equation \ref{eq:barPLE} with additional quadratic or higher order terms in $\log(1+z)$ \citep[e.g.][]{Page97,Silverman08}. Alternatively, \citet{Ueda03} described a PLE model which has 2 different slopes below and above a critical redshift, $z_c$. This parameterisation allows the luminosity evolution to flatten off, cease completely, or increase or decline at high redshifts. We have adopted this form of luminosity evolution, additionally allowing for a smooth transition between the 2 slopes, thus the evolution of $L_*$ is given by
\begin{equation}
\log L_*(z) = \log L_0 - \log\left[ \left(\frac{1+z_c}{1+z}\right)^{p_1} + \left(\frac{1+z_c}{1+z}\right)^{p_2} \right]
\label{eq:PLEdpl}
\end{equation}
where the $z_c$ parameter controls the transition from the strong low-$z$ evolution to the high-$z$ form.

\subsubsection{Luminosity and Density Evolution (LADE)}

The work of \citet{Aird08} indicated that while the XLF retained a similar shape at high redshift, there was an overall decline in the number density at the faint end ($\Lx\lesssim L_*$), yet little or positive evolution at the bright end. This behaviour cannot be described by PLE. We have modified the PLE form described above to additionally allow for some overall decreasing density evolution with redshift, by allowing the normalisation constant $K_\mathrm{norm}$ to evolve as
\begin{equation}
\log K_\mathrm{norm}(z)= \log K_0 + d(1+z)
\end{equation}
thus introducing an additional parameter, $d$. 
With this parameterisation the shape of the XLF stays fixed at all redshifts, thus our evolution consists of an XLF which shifts in both luminosity and overall density. We refer to this evolutionary model as Luminosity And Density Evolution (LADE). This model is similar to the Independent Luminosity and Density Evolution (ILDE) described by \citet{Yencho09}, but with a different form for the density evolution and a more complex luminosity evolution that can flatten off at higher redshifts. 

\subsubsection{Luminosity-Dependent Density Evolution (LDDE)}

As discussed in section \ref{sec:intro}, a number of authors have proposed that an evolutionary scheme in which the XLF undergoes a density evolution which is dependent on luminosity is required to accurately describe the evolution of both soft X-ray selected \citep[e.g.][]{Miyaji00, Hasinger05} and hard X-ray selected \citep[e.g.][]{Ueda03,Silverman08} AGN. We have adopted the parameterisation given by \citet{Ueda03}, in which a typical double power-law form (e.g. equation \ref{eq:lf}) of the XLF at $z=0$ is modified by an evolution term which is a function of luminosity and redshift,
\begin{equation}
\psi(\Lx,z)= \psi(\Lx,0) \; e(\Lx,z)
\end{equation}
where $e(\Lx,z)$ is a power-law function of $z$, which changes between 2 different forms at a cut-off redshift, $z_c$, which depends on the luminosity. Thus
\begin{equation}
e(z,\Lx)=\left\{ \begin{array}{l l} (1+z)^{e_1} & [z<z_c(\Lx)]\\
							e(z_c)\left(\frac{\displaystyle 1+z}{\displaystyle 1+z_c(\Lx)}\right)^{e_2} & [z\ge z_c(\Lx)]\\
							\end{array}\right. 
\end{equation}
where
\begin{equation}
z_c(\Lx)=\left\{ \begin{array}{l l} z_{c*} & [\Lx\ge L_a]\\							 z_{c*}\left(\frac{\displaystyle \Lx}{\displaystyle L_a}\right)^\alpha  & [\Lx<L_a]\\
				\end{array}\right..
\end{equation}
We allow all the parameters to vary in our fitting (although apply priors over constant ranges), thus the evidence, $\mathcal{Z}$, will fully account for the complexity of this parameterisation. 

\begin{table*}
\caption{Best-fit parameters and relative evidence or our different evolutionary models using the hard X-ray selected sample only}
\begin{tabular}{l r@{.}l r@{.}l r@{.}l r@{.}l r@{.}l}
\hline
Parameter &  \multicolumn{2}{c}{Lower limit}$^{a}$ &  \multicolumn{2}{c}{Upper limit}$^{a}$ & \multicolumn{2}{c}{PLE}     & \multicolumn{2}{c}{LADE}    &  \multicolumn{2}{c}{LDDE} \\
\hline
$\log K_\mathrm{norm}$ (or $K_0$) $/$ Mpc$^{-3}$ & -7&0 & -3&0  &  $-4$&$95^{+  0.08}_{-  0.08}$	& $ -4$&$02^{+  0.15}_{-  0.15} $   & $-5$&$91^{+  0.19}_{-  0.19} $\\
$\log L_*$ (or $L_0$) / \ergs			 & 43&0 & 46&0  &  $44$&$82^{+  0.06}_{-  0.06}$	        & $ 44$&$59^{+  0.12}_{-  0.12} $   & $44$&$24^{+  0.11}_{-  0.11} $\\
$\gamma_1$		      			 & -0&1 & 1&5   &  $ 0$&$70^{+  0.03}_{-  0.03}$	& $  0$&$58^{+  0.04}_{-  0.04} $   & $ 0$&$80^{+  0.03}_{-  0.03} $\\
$\gamma_2$					 &  1&5 &  4&0  &  $ 2$&$53^{+  0.10}_{-  0.10}$	& $  2$&$55^{+  0.12}_{-  0.12} $   & $ 2$&$36^{+  0.15}_{-  0.15} $\\
$p_1$						 &  3&0 & 10&0  &  $ 5$&$16^{+  0.35}_{-  0.35}$        & $  6$&$60^{+  0.58}_{-  0.58} $   &  \multicolumn{2}{c}{...}   \\
$p_2$						 & -4&0 & 3&0   &  $-1$&$38^{+  0.22}_{-  0.22}$         & $  0$&$63^{+  0.42}_{-  0.42} $   &  \multicolumn{2}{c}{...}   \\
								   
$e_1$					         & 2&0    & 6&0        &\multicolumn{2}{c}{...}         & \multicolumn{2}{c}{...}          & $  4$&$48^{+  0.30}_{-  0.30} $\\
$e_2$						 & -5&0   & 0&0        &\multicolumn{2}{c}{...}         & \multicolumn{2}{c}{...}          & $ -2$&$85^{+  0.24}_{-  0.24}$\\

$z_c$ (or $z_{c*}$)				 & 0&4     & 3&0  & $ 0$&$89^{+  0.08}_{-  0.08}  $      & $ 0$&$75^{+  0.14}_{-  0.14}$    &  $1$&$89^{+  0.14}_{-  0.14}$\\
$d$						 &-1&5     & 0&0  &  \multicolumn{2}{c}{...}            & $-0$&$36^{+  0.03}_{-  0.03}$    & \multicolumn{2}{c}{...}\\
$\log L_a /$ \ergs 				 & 44&0  & 46&0&     \multicolumn{2}{c}{...}            &  \multicolumn{2}{c}{...}         & $45$&$24^{+  0.19}_{-  0.19}$\\
$\alpha$					 & 0&0     & 1&0  &  \multicolumn{2}{c}{...}            &  \multicolumn{2}{c}{...}         &  $0$&$15^{+  0.01}_{-  0.01}$\\[10pt]
$\Delta \ln \mathcal{Z}$			 &    &        &  &        &  $0$&$0\pm0.3$                           &  $+20$&$3\pm0.3$   & $+ 32$&$9\pm 0.3$   \\
\hline
\end{tabular}\\
\begin{flushleft}
$^{a}$Upper and lower limits on the constant priors for the parameters.
\end{flushleft}
\label{tab:xlf_evol}
\end{table*}
%
\subsection{Results}
\label{sec:results}

Table \ref{tab:xlf_evol} gives the best-fit (posterior mode) parameters for our three evolutionary models using the hard X-ray selected samples only over all redshifts, and values of the Bayesian evidence relative to the evidence for the PLE model, $\Delta \ln \mathcal{Z}$. These values indicate the the LADE model, introducing an additional parameter, provides a significantly better fit than the PLE model. 
There is also very strong evidence favouring the more complicated LDDE parameterisation over either the LADE or PLE models. Figure \ref{fig:LDDE} plots the maximum-likelihood LDDE model along with binned estimates for a range of redshifts. It immediately apparent that our LDDE model predicts a much milder evolution of the faint end of the XLF than the model presented by \citet{Silverman08}, with significantly less flattening of the faint-end slope. This is most likely due to our inclusion of objects with fainter optical magnitudes and correction for remaining incompleteness in the samples (see further discussion in section \ref{sec:evol_comp} below. However, the majority of the flattening of the faint-end slope we do observe is found at $z\gtrsim 1.2$, precisely where our photometric redshifts become unreliable and prone to catastrophic failure, and thus this measurement of the faint end of the XLF will be biased. Therefore, we are unable to confidently rule out the PLE or LADE evolutionary models using the results from direct hard X-ray selection only. 

In table \ref{tab:evol_colsel} we present the results for the three evolutionary models using the colour pre-selected samples at high redshifts in addition to the hard X-ray samples at $z<1.2$. The colour pre-selection approach restricts the sample to sources for which we can reliably estimate the redshift. This should therefore provide more robust measurements of the XLF at high-redshifts that are not subject to systematic biases due to catastrophic failure of photometric redshifts. The sample size is smaller and incomplete, but due to the well-defined selection functions we are able to correct for this incompleteness.
We find strong evidence favouring the LADE or LDDE models over the PLE model. We can thus rule out this simplest model, consisting of an XLF which only evolves in luminosity without changing shape or moving in overall density, and is thus unable to reproduce the lack of evolution of the bright end of the XLF at high redshifts and the observed decline at fainter luminosities. However, there is only very weak evidence ($\Delta \ln \mathcal{Z}=+1.1$) favouring the most complex LDDE evolutionary scheme over the simpler LADE model in which the XLF retains the same shape at all redshifts. We therefore conclude that there is no significant evidence that an LDDE parameterisation, and the associated flattening of the faint-end slope, is required to describe the evolution of the XLF. Our LADE model, in which the XLF retains the same shape but shifts in both luminosity and density, is also able to adequately describe the observed evolutionary behaviour. Our best-fit LADE model indicates strong luminosity evolution takes place between $z=0$ and $z=1$, but is consistent with a value of $p_2=0$, indicating that the evolution of $L_*$ flattens off at high redshifts, and the evolutionary behaviour becomes dominated by the overall negative density evolution. 

Our LADE model is plotted in Figure \ref{fig:xlf_colsel}, evaluated at the centre of a number of redshift bins, along with binned estimates (see section \ref{sec:binned}) using both the hard X-ray samples and the high-$z$ colour pre-selected samples. The LADE model is in good agreement with the bright end of the XLF determined from the hard X-ray selected samples at $z>1.2$ (open circles, not used to constrain this model), but generally predicts a higher number density below $L_*$, reflecting the systematic failure of the photometric redshifts in this regime. At $2.0<z<2.5$ our binned estimates from the colour pre-selection are systematically below the LADE best fit (although agree within $\sim 1\sigma$). Indeed, the binned data appear to indicate that the overall number density drops between \z1 and \z2, but then rises again between \z2 and \z3 (none of our evolutionary models could describe such behaviour). However, it is more likely that this behaviour reflects additional incompleteness in the BX selection, which has not been accounted for. The issues and limitations of our methodology are discussed further in section \ref{sec:limitations} below. There is much better agreement with the $2.5<z<3.5$ binned estimates, where the sample is larger and the XLF is more robustly determined. 

\begin{figure*}
\begin{center}
\includegraphics[width=\textwidth]{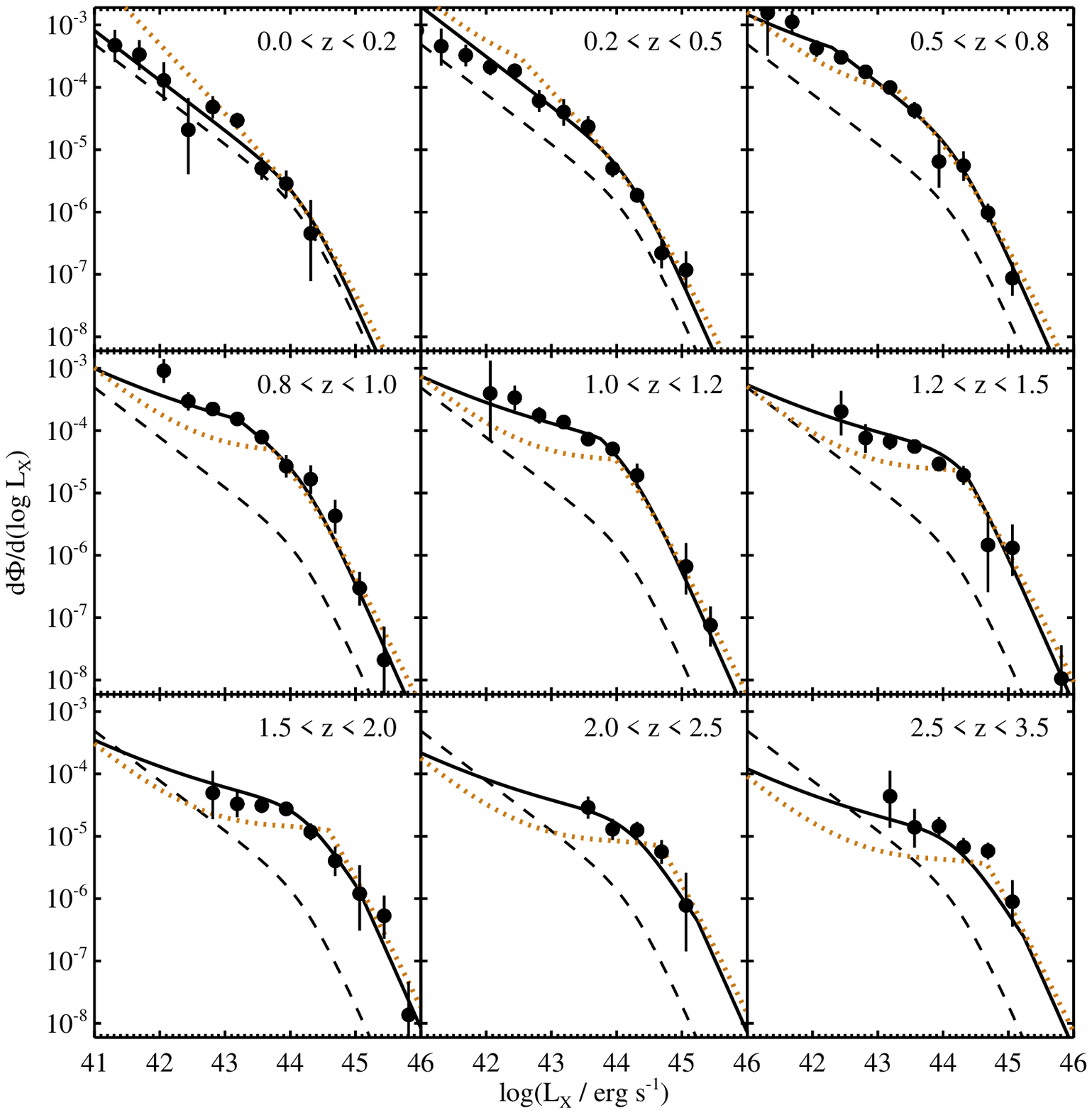}
\end{center}
\caption[Luminosity-Dependent Density Evolution results compared to previous studies]{
Our best fit Luminosity-Dependent Density Evolution model of the XLF using the hard X-ray selected samples only, evaluated at the centres of the displayed redshift bins ({\it solid lines}), along with binned estimates ({\it circles}, see section \ref{sec:binned}). The LDDE model evaluated at $z=0$ is shown in every panel ({\it dashed lines}). We compare our results to the LDDE model presented by \citet[{\it dotted lines}]{Silverman08}, which exhibits significantly greater flattening of the faint-end slope at high redshifts. However, we also note that our model fit will be biased at $z\sim1.2-3$ where our photometric redshift estimates are prone to catastrophic failure.
}
\label{fig:LDDE}
\end{figure*}

\begin{figure*}
\begin{center}
\includegraphics[angle=0, width=\textwidth]{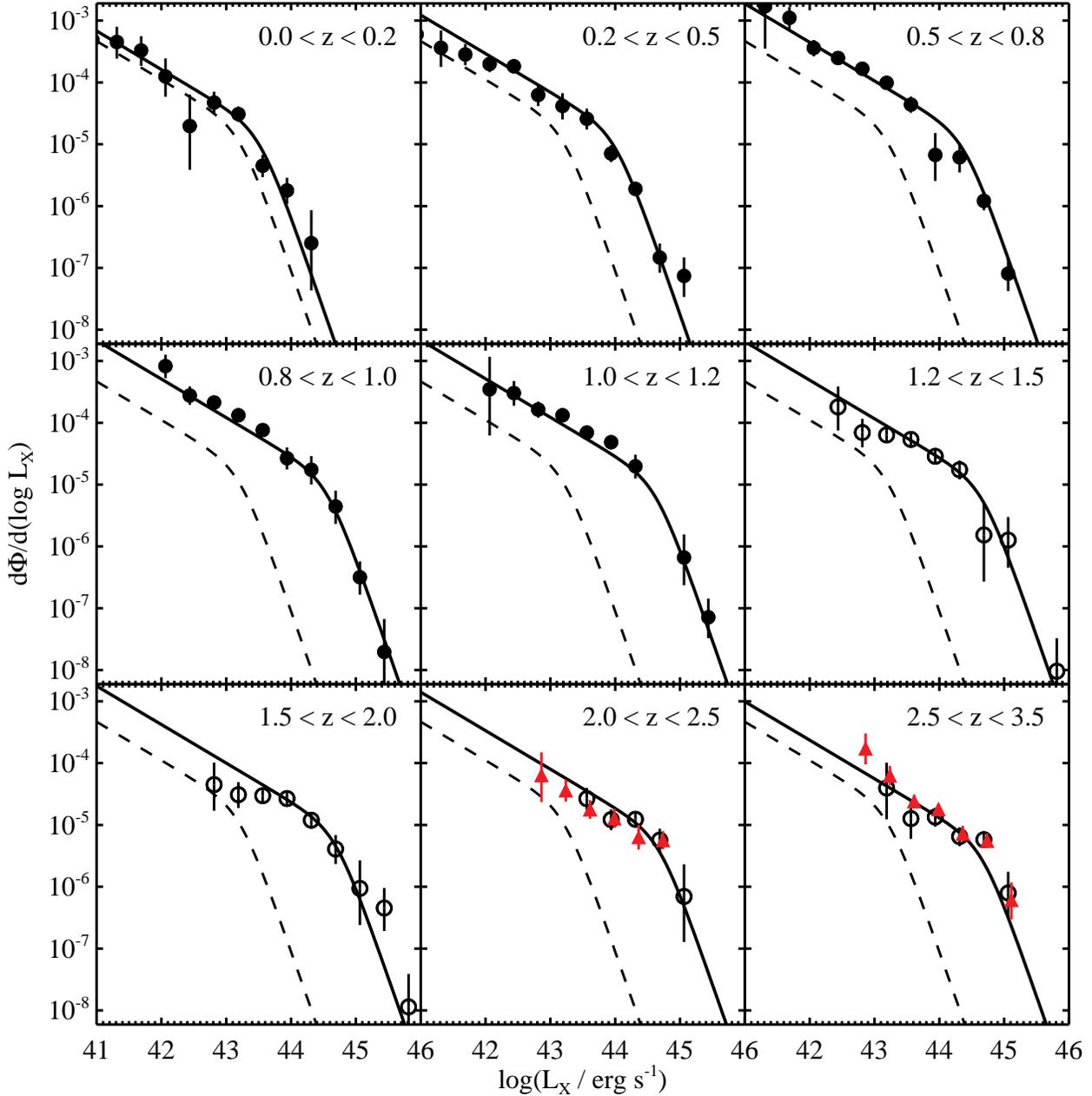}
\end{center}
\caption[XLF results for the LADE model, constrained at high redshifts using the colour pre-selected samples.]{
XLF results for the LADE model, constrained at high redshifts using the colour pre-selected samples as described in section \ref{sec:evol_colsel}. Binned estimates at $z<1.2$ use the hard-band selected sample only, including photometric redshift information ({\it black circles}).  In the final 2 panels binned estimates based on the high-redshift colour pre-selected samples that are used to constrain the evolution are shown (\textit{red triangles}). 
At $z>1.2$ we also show binned estimates using the hard-band selected sample, although these data were not used to fit the model (\textit{open circles}).
The LADE model evaluated at $z=0$ is shown in each panel ({\it black dashed lines}). 
}
\label{fig:xlf_colsel}
\end{figure*}

\begin{table*}
\caption[Best-fit parameters for evolutionary models constrained at high redshifts using the colour selected samples]{
Best-fit parameters and relative evidence for our evolutionary models using the hard-band selected samples at $z<1.2$ and constrained at high redshifts using the colour pre-selected samples. 
}
\begin{center}
\begin{tabular}{l r@{.}l r@{.}l r@{.}l}
\hline
Parameter & \multicolumn{2}{c}{PLE}      &  \multicolumn{2}{c}{LADE}     &  \multicolumn{2}{c}{LDDE}\\
\hline

$\log K_\mathrm{norm}$ (or $K_0$) $/$ Mpc$^{-3}$ & $ -5$&$10^{+  0.04}_{-  0.04} $    & $-4$&$53^{+  0.07}_{-  0.07} $  & $-6$&$08^{+  0.06}_{-  0.06} $\\
$\log L_*$ (or $L_0$) / \ergs		       	 & $ 44$&$96^{+  0.03}_{-  0.03} $    & $44$&$77^{+  0.06}_{-  0.06} $  & $44$&$42^{+  0.04}_{-  0.04} $\\
$\gamma_1$					 & $  0$&$70^{+  0.02}_{-  0.02} $    & $ 0$&$62^{+  0.02}_{-  0.02} $  & $ 0$&$77^{+  0.01}_{-  0.01} $\\
$\gamma_2$				         & $  3$&$14^{+  0.13}_{-  0.13} $    & $ 3$&$01^{+  0.11}_{-  0.11} $  &  $2$&$80^{+  0.12}_{-  0.12} $\\
$p_1$					         & $  5$&$55^{+  0.20}_{-  0.20} $    & $ 6$&$36^{+  0.40}_{-  0.40} $  &  \multicolumn{2}{c}{...}\\
$p_2$					         & $ -1$&$38^{+  0.14}_{-  0.14} $    & $-0$&$24^{+  0.27}_{-  0.27} $  &  \multicolumn{2}{c}{...}\\
$e_1$                                            & \multicolumn{2}{c}{...}          &   \multicolumn{2}{c}{...}         &  $4$&$64^{+  0.24}_{-  0.24}$ \\
$e_2$						 & \multicolumn{2}{c}{...}          &   \multicolumn{2}{c}{...}         & $-1$&$69^{+  0.12}_{-  0.12} $\\
$z_c$ (or $z_{c*}$)				 &  $  0$&$84^{+  0.04}_{-  0.04}$  & $ 0$&$75^{+  0.09}_{-  0.09}$     &  $1$&$27^{+  0.07}_{-  0.07} $\\
$d$						 &  \multicolumn{2}{c}{...}         & $-0$&$19^{+  0.02}_{-  0.02}$     &   \multicolumn{2}{c}{...} \\
$\log L_a /$ \ergs 				 &  \multicolumn{2}{c}{...}         & \multicolumn{2}{c}{...}           & $44$&$70^{+  0.12}_{-  0.12} $\\
$\alpha$		 			 & \multicolumn{2}{c}{...}          & \multicolumn{2}{c}{...}           &  $0$&$11^{+  0.01}_{-  0.01} $\\[10pt]

$\Delta \ln \mathcal{Z}$			 &  $0$&$00\pm0.3$                       &  $+11$&$1\pm0.3   $              & $+12$&$2\pm 0.3$   \\

\hline
\end{tabular}
\end{center}
\label{tab:evol_colsel}
\end{table*}

%
\section{Discussion}
\label{sec:discuss}

Our results for the evolution of the faint end of the 2--10 keV XLF should be the most accurate and robust to date. Our work utilises both of the \chandra\ deep fields, including the full 2 Ms exposure available for each field. The additional large area of deep (200 ks) X-ray data provided by the AEGIS survey are also essential for an accurate determination of the faint-end slope. Our sensitive point source detection techniques allow us to probe to the greatest depths in these data, yet our Bayesian techniques allow us to fully account for uncertainty in the flux measurements at these faint limits, and account for the Eddington bias.  
By incorporating photometric redshift estimates we are able to achieve high redshift completeness ($\sim 75$\%), but we have accounted for the uncertainty in such redshift determinations. We have corrected for remaining incompleteness in our samples using the relation between optical and X-ray fluxes, which we have constrained based on our observed samples. At high redshifts we use colour pre-selection to constrain the XLF and reduce biases due to the catastrophic failures of photo-$z$ estimates. We have also used a likelihood ratio matching technique to ensure we only include robust optical counterparts to X-ray sources.

\subsection{Comparison with previous results}
\label{sec:evol_comp}

Our work offers a number of improvements over previous studies. No previous work has considered the errors in  X-ray fluxes and detection threshold, and thus will not account for Eddington bias. Additionally, no prior studies utilising photometric redshifts \citep[e.g.][]{Barger05,Silverman08,Ebrero09} have accounted for the known uncertainties in such estimates, or the possible effects of catastrophic failures which we believe led to a biased result at high redshifts in our work (see Figure \ref{fig:LDDE}, section \ref{sec:results}). 
The commonly used photometric redshift catalogue for the CDF-N, presented by \citet{Barger03c}, used the same optical data as our work, and the BPZ code, and thus these estimates will be subject to the same uncertainties and potential for catastrophic failures. 
\citet{Zheng04} used a range of photo-$z$ codes, incorporated a wider range of optical filters, deep {\it HST} imaging data, and used the X-ray data as an indication of absorption properties, to provide improved photometric redshift estimates of X-ray sources in the CDF-S (1 Ms data). Although these estimates were found to be accurate to within $\sim 8$\%, the authors discussed the remaining uncertainties, particularly for the fainter sources, and the possibility of catastrophic failure. The lack of spectroscopic follow-up makes it difficult to assess the reliability of all photometric redshifts at high-$z$, and the uncertainties should be accounted for in estimates of the XLF. 

Our completeness corrections allow us to account for the lack of redshift information for a fraction of our sources, which is particularly important for determining the faint end of the XLF, where the optical counterparts are often too faint to be detected in even the deepest available imaging. Previous authors have often omitted any form of completeness correction, instead setting high X-ray flux limits \citep[e.g.][]{Ueda03} or presenting maximal XLFs assigning all unidentified sources to a given redshift bin \citep[e.g.][]{Barger05}.
\citet{Silverman08} did apply a completeness correction, based on the fraction of sources with a given X-ray flux with redshift information, although samples were restricted to objects with bright optical counterparts ($r'<24$) which may introduce additional biases. \citet{Ebrero09} restricted their analysis to samples with high-redshift identification completeness, and applied small additional corrections based on the fraction of identified sources as a function of X-ray flux, although the completeness was likely to be over estimated in the CDF-S as they did not consider the high fraction of spurious counterparts. A similar approach was adopted by \citet{Yencho09}, although their samples were restricted to spectroscopic identifications and thus these completeness corrections were large ($\sim 60$\%) at the faintest flux levels, and may be subject to additional biases and inaccuracies given the large number of complex factors involved in spectroscopic success rates. We thus believe our approach of considering only robust optical counterparts to X-ray sources and modelling the distribution of $f_\mathrm{X}/f_\mathrm{opt}$ ratios, in addition to our colour pre-selection approach at high-redshifts, constitutes a more robust treatment of completeness than in previous work.

In Figure \ref{fig:sdens} we directly compare the evolution of the space density of AGN predicted by our LADE model and a number of recent studies of the evolution of the XLF, all of which concluded that LDDE was taking place.  Results are shown in two luminosity bins, which approximately correspond to above and below the characteristic break luminosity, $L_*$, although $L_*$ does evolve and thus such space density plots may provide a distorted view of the evolutionary behaviour of the XLF. Indeed, in the higher luminosity bin ($44.5<\log \Lx<46$) our model indicates a very different evolution of the space density at $z\lesssim 1$ to previous (LDDE) results, but this is partly due to small offsets in luminosity space which manifest as very large differences in density because of the steep slope of the bright end of the XLF. This can also lead to large differences in binned estimates depending on the methodology or assumed evolution of the XLF. However, in this regime, the XLF remains poorly constrained, and despite the very different evolutionary behaviour our binned results are consistent with the \citet{Silverman08} and \citet{Ebrero09} models at the $\sim 1-2\sigma$ level, although the \citet{Yencho09} model is significantly lower. The high redshift form of our evolution in this luminosity bin predicts similar space densities to the previous studies, although the form of the evolution and location of the peak in the number density is somewhat different. 

Our results are in good agreement at low redshifts ($z\lesssim1$) in the lower luminosity bin ($43<\log \Lx<44.5$), although the \citet{Yencho09} results remain below our estimates. However, at higher redshifts our model predicts a much higher space density than found by \citet{Yencho09} or \citet{Silverman08}. We attribute the decline found by these authors to incompleteness, catastrophic failures of photometric redshift estimates, and false associations of X-ray sources with lower redshift optical counterparts. Conversely, the \citet{Ebrero09} LDDE model is in very good agreement with our result, most likely because of the higher completeness of their samples. However, our results have shown that an LDDE model is not necessary to describe this evolution.

It is worth noting that our LADE form of the evolution is able to describe a ``cosmic-downsizing" behaviour in which the peak in the space density moves to lower redshifts for lower luminosity AGN, and indeed we do find small redshift offsets between the peak space densities for different luminosity ranges (e.g. Figure \ref{fig:sdens}). However, our model does not find the strongly luminosity-dependent shift in peak space density indicated by some prior studies \citep[e.g.][]{Hasinger05,Silverman08}. This is mainly due to our steeper faint-end slope (and thus higher number density of low-luminosity AGN at $z\gtrsim 1$), but is also due to the differences in the form of the bright-end evolution discussed above.

\begin{figure}
\includegraphics[width=\columnwidth]{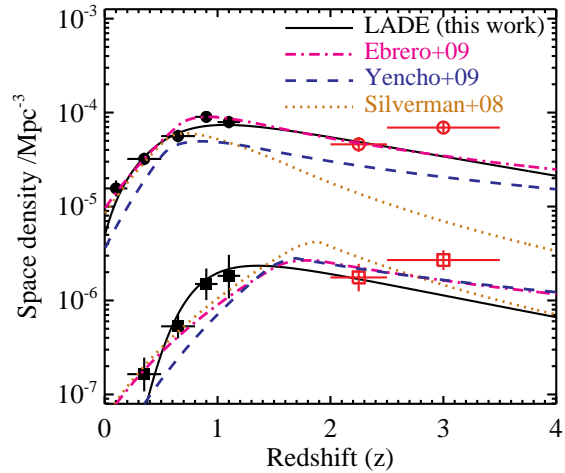}
\caption{
The evolution of the space density of AGN based on our LADE model for two luminosity ranges: $43<\log \Lx<44.5$ (\textit{circles}) and $44.5<\log \Lx < 46$ (\textit{squares}). Binned estimates are shown from the hard X-ray samples at $z<1.2$ (\textit{solid symbols}) and the colour pre-selected samples at \z2--3 (\textit{open symbols}). The space density based on LDDE models determined by \citet[\textit{dot-dashed line}]{Ebrero09}, \citet[\textit{dashed line}]{Yencho09} and \citet[\textit{dotted line}]{Silverman08} are also shown.
}
\label{fig:sdens}
\end{figure}

\subsection{Remaining uncertainties}
\label{sec:limitations}

While we believe our work offers a number of improvements over previous studies, there are remaining uncertainties associated with our approach. 

\subsubsection{Photometric redshifts}
As previously discussed, the introduction of photometric redshifts, essential to improve the completeness of samples at faint X-ray fluxes, does introduce significant uncertainty, which we have tried to account for by considering probability distributions for the redshift estimates. However, there are a number of limitations to photometric redshift techniques that are thus also limitations in our studies. This is particularly apparent in the catastrophic failure of our photo-$z$ estimates at $z\gtrsim1.2$, which we believe led to a biased result for the evolution of the XLF, favouring an LDDE scheme with a flattening faint-end slope (section \ref{sec:results}, table \ref{tab:xlf_evol}). Additionally, the majority of our photo-$z$ estimates are based on template fitting approaches that utilise a variety of galaxy templates (and priors based on the galaxy population), and thus the derived redshift probability distributions may not correspond to the true redshift for an AGN, which could bias our results at low as well as high redshifts. Incorporating AGN templates and using priors based on the AGN population may offer significant improvements and reduce systematic biases and catastrophic failures \citep[e.g.][]{Zheng04,Salvato08}. Expanding the number and wavelength range of the filters, such as including near- and far-IR data, also has potential to improve photometric redshift estimates and provide better fits to AGN templates \citep[e.g][]{Salvato08}. 

In addition to the template fitting approach, we also utilised ANN$z$ redshift estimates in the AEGIS field, in which an artificial neural network is trained to provide photometric redshifts using a sample of objects with spectroscopy. While this approach produced reliable results, it is limited to the subset of objects with a certain range of colours and magnitudes for which follow-up spectroscopy is available. Indeed, spectroscopic validation of photo-$z$ estimates for large numbers of optically-faint X-ray sources is impossible with current instrumentation, which is a major limitation of photometric redshift techniques. 

\subsubsection{High-$z$ colour pre-selection}

To reduce the bias at high redshifts we incorporated rest-frame UV colour pre-selected samples. This should avoid the systematic biases associated with catastrophic failure of the photometric redshift estimates, but does risk introducing additional uncertainty and biases. As noted in section \ref{sec:results}, the binned estimates based on the colour pre-selected samples at $2.0<z<2.5$ (dominated by BX selected objects) fell systematically below the model, ostensibly indicating a rise in the space density of AGN between \z2 and \z3 (also seen in Figure \ref{fig:sdens}). The BX criteria select objects in a very thin slice of colour space, identifying objects with a specific spectral shape rather than strong spectral features.
This makes the BX selection both more susceptible to contamination by lower redshift sources and more prone to incompleteness than the more widely-used LBG selection that identifies a strong break in the spectrum shortward of the Lyman-limit. The shape and efficiency of the derived selection functions will be highly dependent on the modelling of the AGN spectra and the scatter in their properties, and it is quite possible that our relatively simple modelling may have led to overestimates of the completeness. Indeed, both the LBG and BX selections require AGN to be bright in the rest-frame UV, and will therefore miss low-luminosity, moderately or heavily obscured AGN residing in red, evolved host galaxies. 
\citet{Aird08} found that LBG selection recovered a number density of low-luminosity AGN that is at least as high as found via direct X-ray selection and follow-up, tentatively indicating that the AGN population at \z3 is dominated by objects with blue colours \citep[cf. \z1,][]{Nandra07b}. However, this may not be true at slightly lower redshifts, resulting in a lower number density at $2.0<z<2.5$ via the BX selection as red galaxies will be missed. 

As discussed in section \ref{sec:lowzinterlopers}, our combination of LBG and X-ray selection to identify \z3 AGN does not suffer from low-redshift contamination, but this is not the case for the BX selection. Without complete spectroscopic follow-up of the X-ray detected BX candidates we cannot be confident that we have excluded all interlopers (although we believe we are able to identify low-redshift star-forming galaxies, see Figure \ref{fig:fxfopt_highz}), and it is unclear how this might affect our derived evolution of the XLF. Extensions of the BX selection technique, which incorporate data at redder wavelengths, may  allow low-redshift interlopers to be excluded, and help identify AGN in redder host galaxies. Indeed, a hybrid of the photometric redshift and colour pre-selection approaches, utilising data from all available wavebands to perform SED fitting and estimate a redshift, but also performing a number of colour cuts to avoid catastrophic failures, and carefully modelling the distribution of AGN spectral properties to determine the incompleteness, may be the optimal approach to determine the high-redshift evolution.  

It is also worth noting that our approach of combining direct hard X-ray selection at $z<1.2$ and colour pre-selection at high redshifts provides no data for $z\approx 1.2-2.0$. Significant evolution may be taking place over this redshift range, particularly at high luminosities \citep[e.g.][]{Silverman05}. Indeed, at high luminosities incompleteness issues should be less severe, although we note that at \z2 we expect $\sim$10\% of AGN with $\Lx=10^{44.5}$ \ergs\ will have optical magnitudes $R>24$, based on our $\fx/f_\mathrm{opt}$ distributions. In this redshift range, photometric redshift estimates are prone to catastrophic failure, and spectroscopic identification can also be difficult even for bright targets. Thus care must be taken to avoid biased results. As found in section \ref{sec:results}, such systematic biases can erroneously indicate the need for a more complex evolutionary model.

\subsubsection{Completeness corrections}

To account for objects with optical magnitudes below our limit of $R=25.5$, we introduced a simple completeness correction assuming the $\log \fx/f_\mathrm{opt}$ distribution can be described by the same Gaussian function at all redshifts (section \ref{sec:completecorr}). This may be an over simplification, as the redshift distribution of optically-faint sources will differ from the brighter population. Indeed, \citet{Mainieri05} found that optically-faint X-ray sources tended to be found at higher redshifts. Additionally, \citet{Koekemoer04} discuss how Extreme X-ray Objects (EXOs) which are X-ray bright but lack optical counterparts, even in very deep \textit{Hubble} ACS imaging, may represent a population of very high-redshift ($z\sim 6-7$) AGN where the redshifted Lyman-break suppresses all emission at optical wavelengths. More sophisticated modelling of the rest-frame UV--optical properties of the AGN population and the effects on observed magnitudes as a function of redshift would enable improvements to our completeness corrections. Furthermore, the shape and evolution of the optical luminosity function of the AGN host galaxies will affect the completeness of our samples , could alter the shape and efficiency of the optical selection functions, and may systematically impact the photo-$z$ estimates.
In addition, we do not currently utilise any information from the optically-unidentified population (their X-ray fluxes, the fact that they lack counterparts to the limits of our imaging), which could place further constraints on the evolution of the XLF.

\subsubsection{Cosmic variance}
Our investigation of the XLF uses an unprecedented large area of deep 200 ks \chandra\ data in the AEGIS field, which supplements the data from the \chandra\ Deep Fields used in previous studies \citep[e.g.][]{Barger05,Silverman08}. As such, our determination of the faint-end evolution should be less susceptible to the effects of cosmic variance than in prior work. However, cosmic variance could introduce additional uncertainty in our XLF determinations (and thus weaken evidence for more complex evolutionary schemes). Indeed, assuming our X-ray detected AGN cluster according to a power-law with scale length $r_0=7.1$ Mpc and slope $\gamma=1.8$ \citep[typical of X-ray AGN at $z\sim1$][]{Coil09}, we predict a fractional RMS variation of $\sim 5-10$\% in the number density of AGN. \amend{This is comparable to our Poissonian uncertainties due to the number of objects in bin, although the errors will be basically covariant across the luminosity bins, and thus will not affect the shape of the XLF.}


\subsubsection{Intrinsic absorption}

We have calculated rest-frame 2--10 keV luminosities for our sources from their hard band fluxes (or soft band for the high-$z$ sample), assuming a power-law with $\Gamma=1.9$.
This approach should correct for absorption effects due to intrinsic column densities of $\mathrm{N_H}\lesssim10^{23}$ cm$^{-2}$ at $z\sim1$. However, the effectiveness of this correction varies with redshift, and this may bias our measurements of the luminosity, and alter the effective sensitivity of our observations. The extent of these effects will depend on the distribution of AGN absorption properties, which may be evolving with redshift or have a luminosity dependence \citep{Ueda03,LaFranca05}. 
Some previous studies \citep{Ueda03,LaFranca05,Ebrero09} instead attempted to correct for the effects of absorption on a source-by-source basis, and directly account for the $\mathrm{N_H}$ distribution in their calculation of the XLF. 
Determining the extent of absorption is difficult for our faint sources, which have few counts and thus little X-ray spectral information. The effects of absorption on our X-ray sensitivity and completeness calculations will be complicated, and should be corrected for a number of potential selection biases . Such effects are in fact likely to introduce considerable additional uncertainty as to the evolution of the faint end of the XLF, and could thus strengthen our assertions of the lack of a requirement for a more complicated LDDE model to describe the evolution of the XLF.  We defer these considerations to future work. 

%
\subsection{The evolution of AGN accretion activity}
\label{sec:conc_evol}

Our determination of the evolution of the XLF presented in this paper sheds new light on how the distribution of AGN accretion activity evolves over the history of the Universe. We conclude that given the currently available data, the evolution of the XLF can be described by a luminosity and density evolutionary scheme, in which the XLF retains the same shape, but shifts in both luminosity and density with redshift. A more complicated luminosity-dependent scheme does not provide a significantly better description of the data; even so our best-fit LDDE for the hard X-ray sample only, which we expect is biased at high redshifts, indicates a steeper faint-end slope and less evolution of the shape of the XLF than most previous work. 

\citet{Hopkins05} presented an interpretation of the AGN luminosity function in which the bright end traces the mass distribution of black-holes accreting at the peak rate over their lifetimes, while the faint end corresponds to AGN during transitionary periods as they approach or decline from their peak rate of activity. Furthermore, \citet{Hopkins06} proposed that a luminosity-dependent lifetime for AGN could explain any observed flattening of the faint-end slope, as less time is spent in the transitionary period by the more luminous AGN found at high redshifts. Our results, however, show that the shape of the XLF does not change, thus a strongly luminosity-dependent lifetime may not be required. The lack of evolution in the shape of the XLF indicates the processes involved in the triggering and feeding of AGN remain the same at all redshifts, but increase in overall density from the earliest times until \z1, and moving to lower luminosity systems between $z\approx 1$ and the present day. Such behaviour could be consistent with a merger-driven model of the build-up of galaxy and black-hole mass, and the triggering of AGN \citep[e.g.][]{Kauffmann00,Hopkins08}, in which the hierarchical build-up of galaxies increases the number density of AGN in the early history of the Universe, but the exhaustion of gas supplies in the most massive galaxies at later times \citep[e.g.][]{Cowie96}
results in the ``down-sizing" of AGN to lower luminosities. Fully reconciling such a picture with the full range of observational data for AGN, in addition to the observed evolution of galaxies, using current theoretical models clearly requires significant further investigation. Indeed a number of other processes may play a role in triggering and fuelling low luminosity AGN, such as minor mergers \citep{Hernquist95}, accretion of gas from a hot halo \citep{Fabian95} or bar-driven accretion \citep{Sellwood99}. It is essential to assess the importance of such processes and understand the role they play in the evolution of AGN accretion activity.
\begin{figure}
\includegraphics[width=\columnwidth]{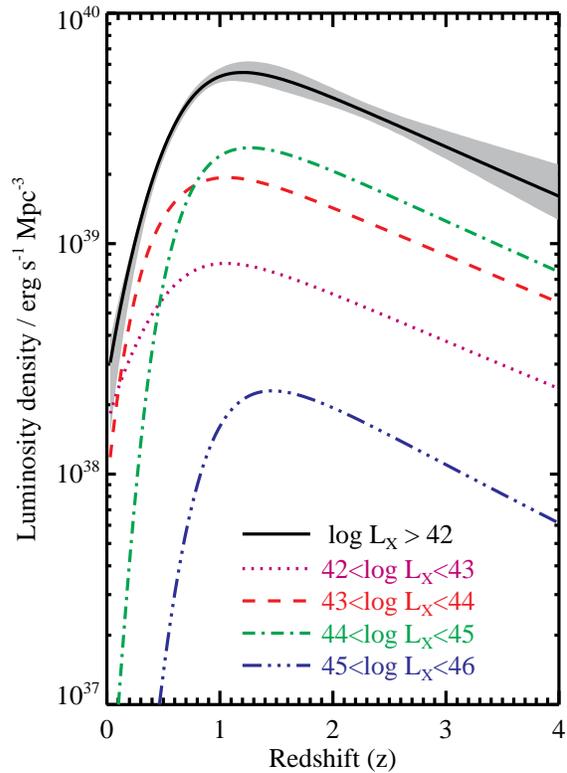}
\caption[Space density and luminosity density of AGN as a function of redshift]{
2--10 keV luminosity density of AGN as a function of redshift for our LADE model integrated over the luminosity ranges indicated. The grey shaded region indicates the 1$\sigma$ uncertainty on the total luminosity density of AGN. The luminosity density is dominated by moderate luminosity ($43<\log \Lx<45$) AGN, and exhibits a mild decline above $z\approx 1.2$.}
\label{fig:ldens}
\end{figure}

In Figure \ref{fig:ldens} we plot the 2--10 keV luminosity density, $\int \Lx \phi(\Lx,z) \dd \log \Lx$, as a function of redshift,  which provides a tracer of the total AGN accretion activity. The evolution of the total luminosity density for all AGN with $\Lx>10^{42}$ \ergs\ is shown as well as the contribution of AGN in set luminosity ranges.
The total luminosity density peaks at $z=1.2\pm0.1$, rapidly declining to lower redshifts, but the decline to higher redshifts is much milder indicating significant AGN activity is taking place out to \z3--4. AGN with luminosities in the range $\Lx\approx 10^{43-45}$ \ergs (i.e.  $\sim L_*$) are responsible for the majority of the luminosity density, although the contribution of $\Lx=10^{44-45}$ \ergs\ AGN falls off rapidly at $z\lesssim 0.8$ as $L_*$ evolves, and $\Lx=10^{43-44}$ \ergs\ AGN dominate. 

We can track the build-up of black-hole mass more directly by relating AGN luminosity to mass accretion, as first proposed by \citet{Soltan82},
\begin{equation}
L_\mathrm{bol}=\epsilon \dot{M_\mathrm{acc}}c^2= \frac{\epsilon \dot{M_\mathrm{bh}}c^2}{1-\epsilon}
\label{eq:macc}
\end{equation}
where $\dot{M_\mathrm{acc}}$ is the mass accretion rate, $\dot{M_\mathrm{bh}}$ is the rate of change of black-hole mass density, $L_\mathrm{bol}$ is the bolometric luminosity, $\epsilon$ is the radiative efficiency of the accretion process and $c$ is the speed of light. We have adopted a simple approach \citep[e.g.][]{Barger05}, converting our 2--10 keV X-ray luminosities to bolometric values using a constant conversion factor \citep[40,][]{Elvis94}, and assuming a single value of the radiative efficiency, $\epsilon=0.1$ \citep{Marconi04}. We note that a number of authors have discussed the need for luminosity-dependent bolometric corrections \citep[e.g.][]{Marconi04,Hopkins07b}, and \citet{Vasudevan07} reported significant object to object variation in bolometric corrections, which may depend on the Eddington ratio. The radiative efficiency may also vary significantly between AGN, depending on the spin of the black-hole \citep[e.g.][]{Thorne74} in addition to the specifics of the accretion processes \citep[e.g.][]{Merloni08}. However our simple assumptions allow an initial investigation of the consequences of our derived XLF evolution on the build-up of black-hole mass. Using equation \ref{eq:macc} we can convert our luminosity densities to a mass accretion rate, and thus calculate the total black-hole mass density as built up by accretion activity over the history of the Universe. Our results are shown in Figure \ref{fig:bh_acc}.

\begin{figure}
\begin{center}
\includegraphics[width=\columnwidth]{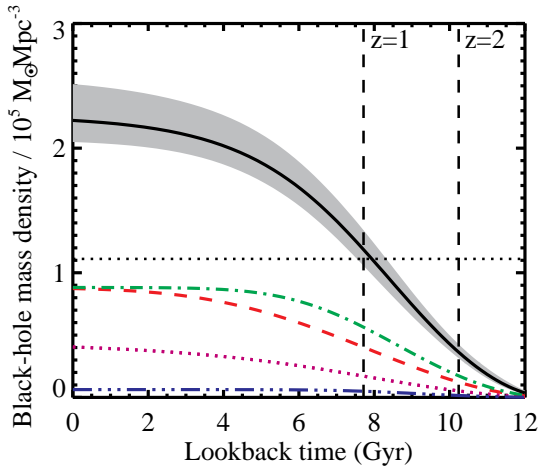}
\end{center}
\caption[Accreted black-hole mass density against lookback time]{
Total accreted black-hole mass density against lookback time based on our LADE ({\it solid}, grey shaded region indicates 1$\sigma$ uncertainty in the derived model). The lower curves correspond to the same luminosity ranges indicated in Figure \ref{fig:ldens}. We find $\sim 50$\% of the local black-hole mass density is built up in AGN actively accreting at $z\gtrsim1$.  
}
\label{fig:bh_acc}
\end{figure}

Based on our LADE model we predict a local black-hole mass density of $2.2\pm0.2 \times 10^5\; M_{\sun}$ Mpc$^{-3}$, where the error reflects the uncertainties in our model fit, but not the potentially larger uncertainties in bolometric correction or accretion efficiency. This value is in good agreement with the estimate of \citet{Yu02} ($2.5\pm0.4 \times 10^5\; M_{\sun}$ Mpc$^{-3}$) based on velocity dispersions of early-type galaxies in SDSS and the $M_\mathrm{bh}-\sigma$ relation, although is lower than the estimate of \citet{Marconi04} ($4.6^{+1.9}_{-1.4} \times 10^5\; M_{\sun}$ Mpc$^{-3}$), possibly indicating that our XLF does not provide a complete census of the history of accretion activity. Figure \ref{fig:bh_acc} shows that a significant fraction ($\sim 50$\%) of this total mass density is accreted at $z\gtrsim 1$. While the majority of the mass build up takes place in moderate luminosity AGN ($\Lx=10^{43-45}$ \ergs), a significant fraction is accumulated at lower luminosities ($\Lx=10^{42-43}$ \ergs). The LDDE model from section \ref{sec:xlf_evol} predicts a lower local black-hole mass density, mainly due to the smaller numbers of AGN at these low luminosities and high redshifts. The redshift range \z1--3 clearly corresponds to a period of significant AGN activity, and thus it is essential to accurately measure the XLF down to $\Lx\approx 10^{42-43}$ \ergs\ in this epoch to determine the history of black-hole mass accretion. 

We can also compare our derived black-hole mass accretion rates to star-formation rates, which we show in Figure \ref{fig:sfr}. Previous studies \citep[e.g.][]{Boyle98b,Silverman08} have shown close similarities between the rapid increase in star-formation rate and black-hole accretion at $z\lesssim 1$, which we confirm. Our results indicate that this correlation continues out to high redshifts \citep[cf.][]{Silverman08}, at least when comparing to the recent star-formation rates of \citet{Bouwens07}, although we note that our model is extrapolated far beyond the redshift range probed by our data. Comparisons of the galaxy and AGN luminosity functions may reveal differences in the details of the evolving distributions of activity, which could reveal further facets of the co-evolutionary processes and the feedback regulating AGN activity and star-formation. 

\begin{figure}
\begin{center}
\includegraphics[width=\columnwidth]{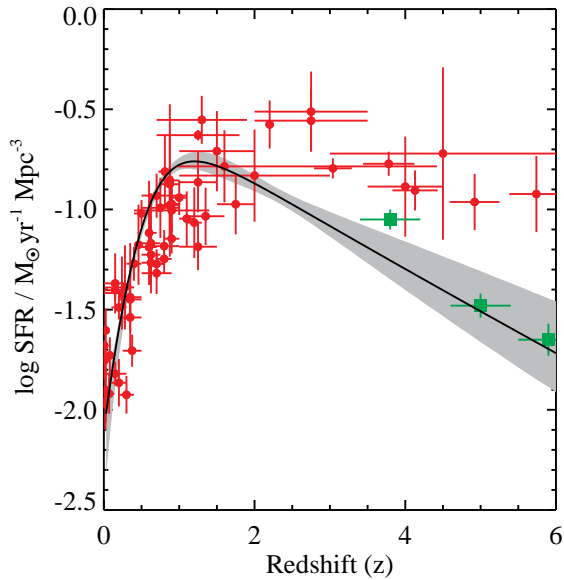}
\end{center}
\caption[Comparison of the star-formation and black-hole accretion rates as a function of redshift]{
Comparison of star-formation rates ({circles}) from the compilation of \citet{Hopkins04} and black-hole accretion rate based on our LADE model ({\it solid line}; grey shading indicates 1$\sigma$ uncertainty in the model). Black-hole mass accretion rate assumes a constant bolometric correction and accretion efficiency, $\epsilon=0.1$, and is scaled up by a factor 5000 \citep{Silverman08}. We also show recent measurements of the star-formation rate at high redshifts from \citet[{\it squares}]{Bouwens07}, which indicate the correlation between star-formation and black-hole accretion rate continues at high redshifts. 
}
\label{fig:sfr}
\end{figure}

\section{Summary}

We have presented new observational determinations of the X-ray luminosity function of AGN. We utilised a hard X-ray selected sample from both the 2 Ms \chandra\ Deep Fields and the large area, deep 200 ks \chandra\ survey in the AEGIS field. A likelihood ratio method was employed to match X-ray sources to optical counterparts and ensure only robust associations were considered. To improve our redshift completeness, we supplemented spectroscopic identifications with photometric redshifts. However, we found our photo-$z$ estimates were systematically biased and prone to catastrophic failure at $z\gtrsim1.2$. We therefore adopted a rest-frame UV colour pre-selection approach to constrain the XLF at high redshifts. By carefully modelling the expected colour distributions of AGN, performing simulations of the optical data and repeating the source recovery procedure, we were able to determine well-defined selection functions that allowed us to correct for the incompleteness associated with the pre-selection technique. 

We developed a sophisticated Bayesian methodology to determine the evolution of the XLF, which accounts for the uncertainties in photometric redshift estimates, the Poissonian nature of the X-ray flux estimate, the fraction of sources with counterparts below the magnitude limits of the optical data, and the optical selection functions at high redshifts. Using a Bayesian model comparison approach, we found that a Pure Luminosity Evolution was unable to adequately describe the evolution of the XLF, but a scheme in which the XLF evolves in both luminosity and density (LADE), but does not change shape, did provide a good fit. We did not find significant evidence for a more complex luminosity-dependent density evolution (LDDE), and the associated flattening of the faint-end slope. However, an LDDE model was required if the direct X-ray selected samples and the photometric redshift estimates were utilised at $z>1.2$, but we believe such a result to be biased due to the inclusion of unreliable photo-$z$ estimates. 

The form of our derived evolution of the XLF differs from many previous results \citep[e.g.][]{Silverman08,Ebrero09,Yencho09}. Our XLF retains the same shape at all redshifts, which may have a number of consequences for the evolution of the black-hole mass function, AGN lifetimes and duty cycles, and the processes that fuel and trigger AGN. 
We find that the total luminosity density peaks at $z=1.2\pm 0.1$. While there is a mild decline at $z>1.2$, significant AGN activity is still taking place at high redshifts.
We find significantly higher number densities of low luminosity AGN ($\Lx \lesssim 10^{44}$ \ergs) at $z\gtrsim1$, which make a significant contribution to the to the total luminosity density and growth of black-hole mass at these redshifts. Our results do not indicate a strong shift of the peak space density to lower redshifts for lower luminosities, although a small shift is predicted by our best-fit evolutionary model. The mild decline in AGN activity to high redshifts appears to correlate with that seen in the star-formation rate, consistent with a co-evolutionary scheme for black-holes and galaxies.

\section*{Acknowledgements}
We acknowledge financial support from STFC (JA, ESL). Some observations reported here were obtained at the MMT Observatory, a 
joint facility of the Smithsonian Institution and the University of Arizona. 
Also based in part on observations obtained with MegaPrime/MegaCam, a joint  
project of CFHT and CEA/DAPNIA, at the Canada-France-Hawaii Telescope  
(CFHT) which is operated by the National Research Council (NRC) of  
Canada, the Institut National des Sciences de l'Univers of the Centre  
National de la Recherche Scientifique (CNRS) of France, and the  
University of Hawaii. This work is based in part on data products  
produced at TERAPIX and the Canadian Astronomy Data Centre as part of  
the Canada-France-Hawaii Telescope Legacy Survey, a collaborative  
project of NRC and CNRS.



\label{lastpage}
\end{document}